\documentclass[12pt,preprint]{aastex}
\usepackage{graphicx}
\usepackage{natbib}
\usepackage{rotating}

\shorttitle{Photometric search for transiting planets from the Italian Alps}
\shortauthors{Damasso et al.}

\begin{document}


\title{Photometric transit search for planets around cool stars from the Western Italian Alps: 
a site characterization study \thanks{Based on observations made at the 
Astronomical Observatory of the Autonomous Region of the Aosta Valley (Italy)}}


\author{M. Damasso\altaffilmark{1,2}, P. Giacobbe\altaffilmark{3}, 
P. Calcidese\altaffilmark{1}, A. Sozzetti\altaffilmark{4}, M.G. Lattanzi\altaffilmark{4}, 
A. Bernagozzi\altaffilmark{1}, E. Bertolini\altaffilmark{1}, and R.L. Smart\altaffilmark{4} 
} 
\altaffiltext{1}{Astronomical Observatory of the Autonomous Region of the Aosta Valley, Loc. Lignan 39, 11020 Nus (Aosta), Italy}
\altaffiltext{2}{Dipartimento di Astronomia, Universit\`a di Padova, Vicolo dell'Osservatorio 5, I-35122 Padova, Italy}
\altaffiltext{3}{Dipartimento di Astronomia, Universit\`a di Trieste, Via Tiepolo 11, I-34143 Trieste, Italy}
\altaffiltext{4}{INAF - Osservatorio Astronomico di Torino, Via Osservatorio 20, I-10025 Pino Torinese, Italy}


\begin{abstract}
We present the results of a site characterization study carried out at 
the Astronomical Observatory of the Autonomous Region of the Aosta Valley (OAVdA), 
in the Western Italian Alps, aimed at establishing its potential to host a photometric 
transit search for small-size planets around a statistically significant 
sample of nearby cool M dwarfs. For the purpose of the site testing campaign, we gathered photometric and seeing 
measurements utilizing different instruments available at the site. 
As in any search for new locations for astronomical observations, we gauged 
site-dependent observing conditions such as night-sky brightness, photometric precision, 
and seeing properties. Public meteorological data were also used in order to help 
in the determination of the actual number of useful observing nights per year.
The measured zenithal $V$-band night-sky brightness is typical of that of very good, 
very dark observing sites. The extinction registered at $V$ band is not dissimilar 
from that of other sites. The median seeing over the period of in situ observations is found to be 
$\sim1.7^{\prime\prime}$. Given the limited duration of the observations, we did not probe 
any possible seeing seasonal patterns, or the details of its possible dependence on other 
meteorological parameters, such as wind speed and direction. Moreover, our data show that 
the seeing at the observatory was reasonably stable during most of the nights. 
The fraction of fully clear nights per year amounts to 39\%, while the total of 
useful nights increases to 57\% assuming a (conservative) cloud cover of not more than 50\% of the night. 
Based on the analysis of photometric data collected over the period May-August 2009 for three stellar 
fields centered on the transiting planet hosts WASP-3, HAT-P-7, and Gliese 436, 
we achieve seeing-independent best-case photometric precision $\sigma_\mathrm{ph}\lesssim3$ 
mmag (rms) in several nights for bright stars ($R\lesssim 11$ mag). 
A median performance $\sigma_\mathrm{ph}\sim6$ mmag during the observing period is obtained for stars with $R\lesssim13$ mag. 
A byproduct of the significant amount of photometric data collected 
in the stellar fields of WASP-3 and HAT-P-7 was the identification of a handful of new 
variable stars, four of which were presented and discussed here. 
Our results demonstrate the OAVdA site is well-poised to conduct an upcoming 
long-term photometric survey for transiting low-mass, small-size planets around a well-defined 
sample of M dwarfs in the solar neighborhood.
\end{abstract}





\section{Introduction}

The discovery by Charbonneau et al. (2009) of a $\sim 6.5$ $M_\oplus$ 
planet (a `Super Earth') with a radius of $\sim 2.7$ $R_\oplus$ in 
transit across the disk of the late-M dwarf GJ 1214 with a period of $\sim1.6$ days 
has heralded the first success for a novel approach to ground-based photometric 
searches for transiting planets. 
Until recently, two main typologies of search programs for planetary transits had been 
adopted: 1) deep, narrow-field surveys using medium- to
large-size telescopes with a narrow field of view (FOV) to monitor many faint sources, 
and 2) shallow, wide-field surveys utilizing small telescopes
with a wide FOV to probe for photometric variability a large number of bright, relatively 
nearby solar-type stars. The former approach has produced so far the overwhelming majority of 
transiting planet detections, and is the preferred choice for ongoing (CoRoT, Kepler) and 
planned (PLATO, TESS) space-borne transit searches (for a review, see Charbonneau et al. 2007). 

Among the many paradigms which the fast-developing field of extrasolar planet science is 
continuously reformulating, a special focus has been given in recent times to the choice of the 
stellar type to be probed for planets. Driven by the available instrument performance for a 
given technique (and by a small amount of hard-to-die Solar-System-centrism), 
initial planet search surveys (be they spectroscopic or photometric) 
focused exclusively on main-sequence, solar-type stars. This is changing, however, with 
increasing awareness of the need to extend the number of detected planets around varied star types, 
for a proper, global understanding of the complex processes of planet formation and evolution. 
With the primary goal of reaching sensitivity to short-period planets with (minimum) masses lower than the 
mass of Neptune, a number of dedicated radial-velocity surveys  have been targeting low-mass stars 
(e.g., Endl et al. 2006; Zechmeister et al. 2009 and references therein). High-precision Doppler data (Rivera et al. 2005; Udry et al. 2007; 
Forveille et al. 2008; Mayor et al. 2009) have unveiled the existence of a population of 
low-mass, close-in planets with $M_p\sin i\approx 2-10$ $M_\oplus$ around cool stars\footnote{Evidence for their presence 
around solar-type stars has been presented by Bouchy et al. 2008; Fischer et al. 2008, 
and more recently by Howard et al. 2009; Vogt et al. 2009; Rivera et al. 2009}. At wider 
separations, Super-Earths around M dwarfs have also been revealed by microlensing surveys (Beaulieu et al. 2006). 

As for the transit technique, the combination of photometric data and Doppler measurements 
for targets at the bottom of the main sequence implies the possibility of deriving estimates 
of the bulk composition for Super-Earth planets (with typical radii in the range $2-4$ $R_\oplus$), 
with requirements on the photometric precision necessary for their detection which are readily fulfilled from the ground 
(for details see for example Nutzman \& Charbonneau 2008). However, the specific choice of targets has 
imposed a re-definition of the observing strategies. Given the sparseness of bright M dwarfs 
throughout the whole sky, the wide-field approach can only be carried out effectively utilizing 
4-m class telescopes, and shifting to near-infrared observing wavelengths in order to be able to 
probe tens of thousands of M dwarfs which would be otherwise too faint in the visible, 
as it is done for example by the WFCAM/UKIRT Transit Survey (WTS, see~\url{http://star.herts.ac.uk/RoPACS/}). 
Alternatively, the narrow-field approach has been reformulated by designing transit 
search programs which aim at probing {\it individually} all the sparsely spread bright, 
nearby M dwarfs in the sky, utilizing a cluster of identical telescopes at one location, 
such as the MEarth project (Nutzman \& Charbonneau 2008). 
The discovery by Charbonneau et al. (2009) is the proof of the 
cunningness of the latter method, given that only for transiting Super-Earth planets orbiting 
nearby small stars it is possible to carry out more in-depth characterization studies, such as 
spectroscopic investigations of their atmospheres.

The ground-breaking discovery of GJ 1214b has, however, also highlighted some of the present limitations 
in our understanding of transiting systems composed of low-mass primaries and small-size planets. 
The first source of uncertainty comes from the existing discrepancies between theory and observations 
in the determination of the sizes of (late) M dwarfs which, as reported by Charbonneau et al. (2009, and 
references therein), are of the order of $10\%-15\%$. This problem, likely stemming from 
the need to introduce in stellar evolutionary models the detailed treatment of the effects 
of non-zero magnetic fields on the properties of low-mass stars (e.g., Mullan \& McDonald 2001; Ribas 2006; 
Torres et al. 2010, and references therein), has significant repercussions on the inferred composition of the planet. 
An important by-product of the long-term photometric monitoring of a large sample of low-mass 
M dwarfs is the discovery of new eclipsing binary systems with low-mass components 
(e.g., Irwin et al. 2009), leading to a sample of accurate measurements of their radii 
which can in turn provide a useful testing ground for improving stellar evolution models. 
Secondly, degeneracies in the models of the physical structures of 
Super-Earths, such as objects with very different compositions having similar masses and radii, 
also prevent one from exactly inferring their interior composition 
(Seager et al. 2007; Rogers \& Seager 2010. See also discussion in Charbonneau et al. 2009) when 
only a mass and radius measurement are available. It is thus clear that more 
discoveries of transiting low-mass, small-radius planets over a range of periods around nearby, 
relatively bright late-type stars are highly desirable. This is of particular relevance as 
the further spectroscopic characterization studies of the atmospheres possible for these planets are one 
of the most effective ways to mitigate the degeneracies of exoplanet interior composition models 
(see for example Miller-Ricci et al. 2009). 

If the wild diversity of transiting giant planets properties is a proxy for what to expect from 
transiting Neptunes and Super Earths, and if the prediction from recent population synthesis 
models of planetary systems formation are on target, which envision increasing planet frequencies 
with decreasing planet mass, one then realizes how well justified is the effort being invested in present 
and planned transit search programs, both from the ground and in space. 
One lesson learned from the experience of wide-field ground-based transit surveys 
for giant planets around bright solar-type stars is that, in order to optimize the phase coverage and 
thus maximize the chances of success, a distributed network of telescopes should be built. Multi-site campaigns carried out 
by the HAT-Net and Super-WASP networks have contributed the large majority of the transiting 
planet discoveries to-date (the HAT-South network is now coming online, see Bakos et al. 2009). 
With this in mind, our group has started investigating the prospects for a Europe-based 
photometric transit search for small-size planets orbiting low-mass stars using a network of 
small telescopes. As a first step towards this goal, we have attempted to gauge the potential of the 
recently inaugurated (2003) Astronomical Observatory of the Autonomous Region of the Aosta Valley (OAVdA,
\url{http://www.oavda.it}) to become the primary node of such a network. Due to its short 
history, OAVdA is significantly involved in teaching and public outreach activities, but it also has 
an early record of collaborative efforts in a variety of scientific programs (e.g., monitoring of 
near-Earth objects and blazar events). The primary reasons for an in-depth characterization 
of the potential of the site for the purpose of a transit search program are as follows. 
First, the Observatory is located in the North-West of the Italian Alps (45.7895 North, 7.47833 East), 
at 1,675 m above sea level, thus its high-altitude location and remoteness provide almost
light-pollution-free night-sky conditions. Second, clear-sky observing conditions are realized at 
OAVdA on $\sim$57$\%$ of the total nights in a year, which is a conservative estimate based on 
4-yr statistics collected at the site (see \S~\ref{nights}). 
Third, the OAVdA site has two platforms with electronically-controlled track-mounted ceilings
that can be completely opened to the sky. One of these can host a variable number of telescopes, 
depending on system specifications and usable area on the platform. 
The availability of a single movable enclosure (the site is essentially unique in Europe) 
and of much of the associated support infrastructure at OAVdA will allow in perspective very 
substantial cost savings for the project. 

The generic design requirements for a ground-based photometric transit search for small-size 
planet around M dwarfs which uses a network of small, identical robotic telescopes 
(optimal bandpass and field of view, telescope aperture, telescope time allocation, observing strategy) 
have been laid out by Nutzman \& Charbonneau (2008). That study assumes that the well-characterized 
Mt. Hopkins site in Arizona chosen for the survey guarantees that a) the limiting system performance (photometric 
precision) for transit detection is achieved, and b) the requested number of useful observing nights 
per year is reached. This work presents the details of a characterization study carried out at OAVdA, which aims at 
demonstrating that the main site-specific requirements (night-sky brightness, number of useful nights, 
seeing, photometric stability) are satisfied, and OAVdA is 
therefore an optimal choice for hosting a precision photometric search for low-mass, 
small-size planets around a well-defined sample of M dwarfs in the solar neighborhood which utilizes 
an array of small telescopes. 

The paper is organized as follows. In \S~2 we describe the instrumentation 
presently available at OAVdA, which was utilized for the purpose of the feasibility 
study, and outline in \S~3 the various steps involved in the dedicated pipeline for the data processing and 
analysis of the photometric data under development. 
We present in \S~4 the observing strategy adopted and the main results of the 
feasibility study in terms of overall site characterization (i.e. the achievable photometric 
precision as a function of atmospheric conditions). We conclude in \S~5 by discussing 
the preparatory steps for a long-term photometric monitoring program to characterize the micro-variability 
characteristics of and search for transiting small-size planetary companions to a well-defined sample of 
low-mass stars, to be carried at the OAVdA site in the near future.

\section{System description}\label{system}

\subsection{Hardware}\label{hardware} 

\begin{figure}[h]
   \centering
   \includegraphics[scale=.6]{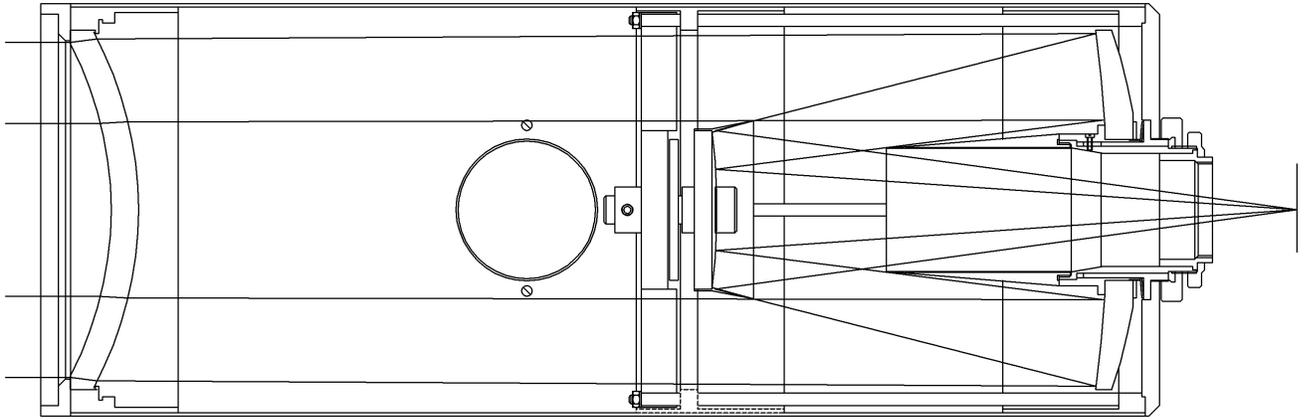}
      \caption{Optical scheme of the Maksutov reflector telescope (aperture 250 mm) 
      used for the characterization study (from http://lnx.costruzioniottichezen.com).}
         \label{optscheme}
\end{figure}

\begin{figure}[h]
   \centering
   \includegraphics[scale=1.]{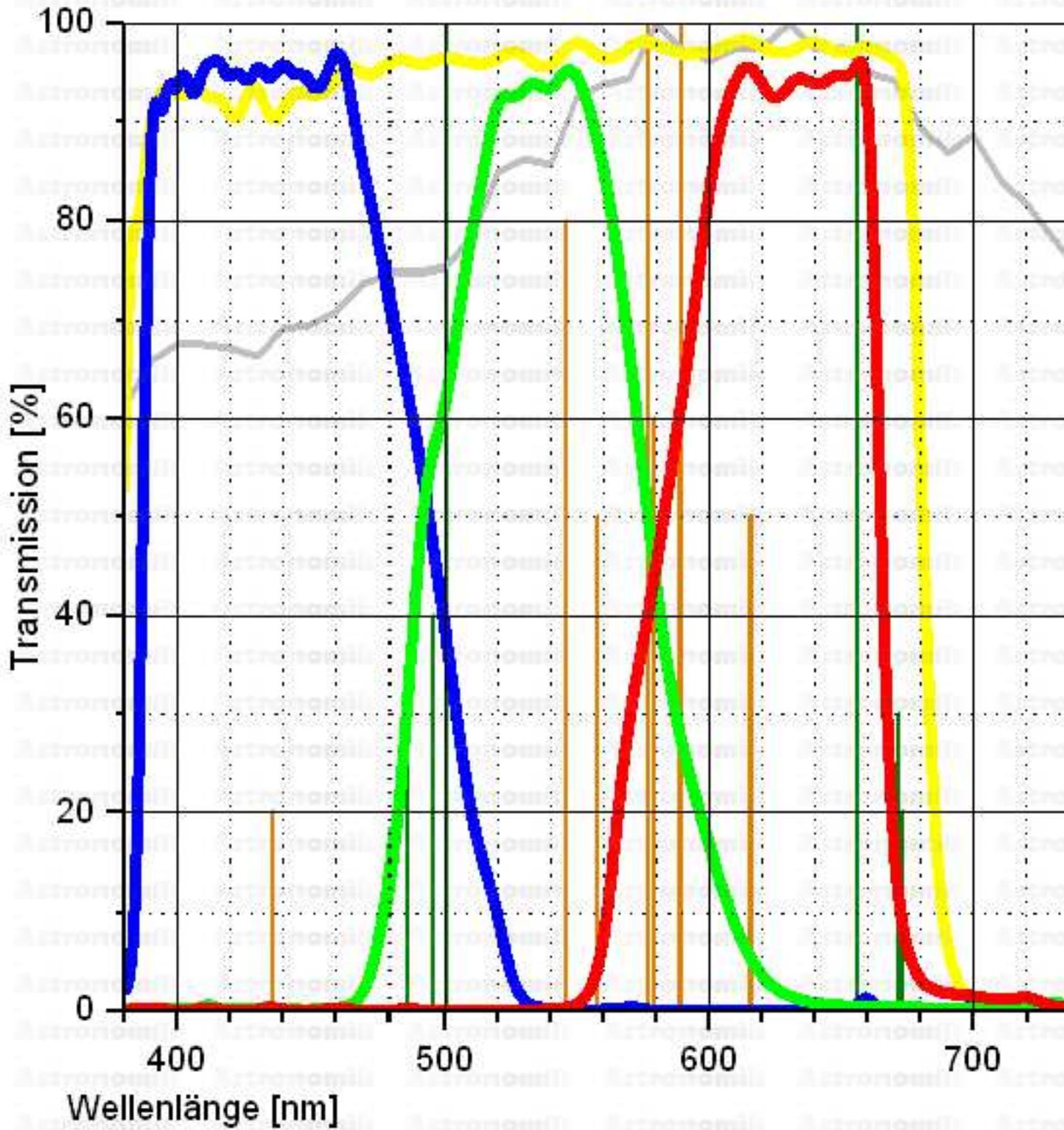}
     \caption{Transmission curve for the set of filters coupled to the CCD in the 250mm telescope. 
      For the site testing we used the $R$ filter. The corresponding transmission curve is showed in red.}
        \label{filters}
\end{figure}

\begin{figure}[h]
   \centering
   \includegraphics[scale=.8]{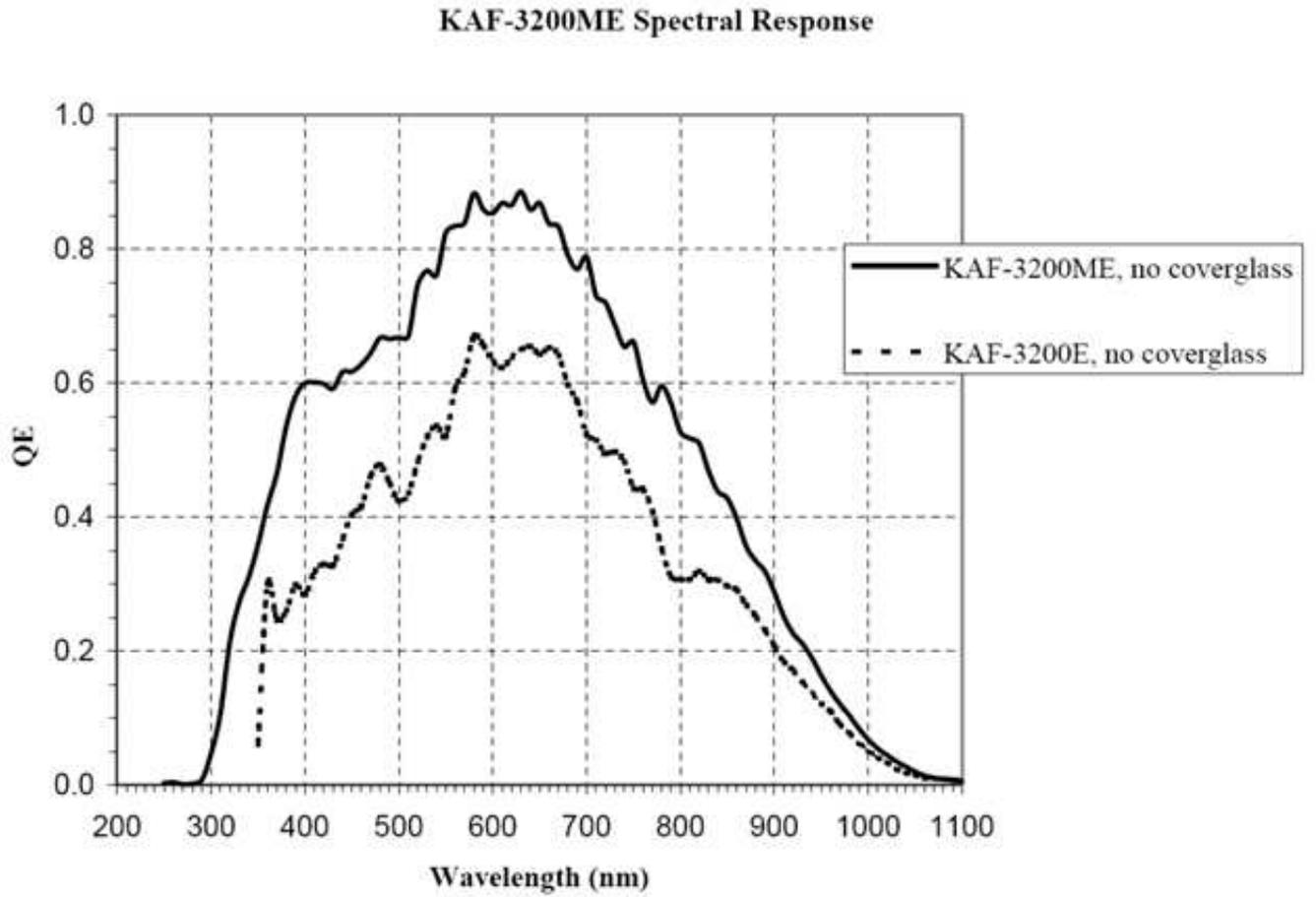}
      \caption{Nominal response curve of the sensor KAF-3200ME integrated in the CCD camera used in this study (solid line).}
         \label{qe}
\end{figure}

\begin{figure}[h]
   \centering
   \includegraphics[scale=.8]{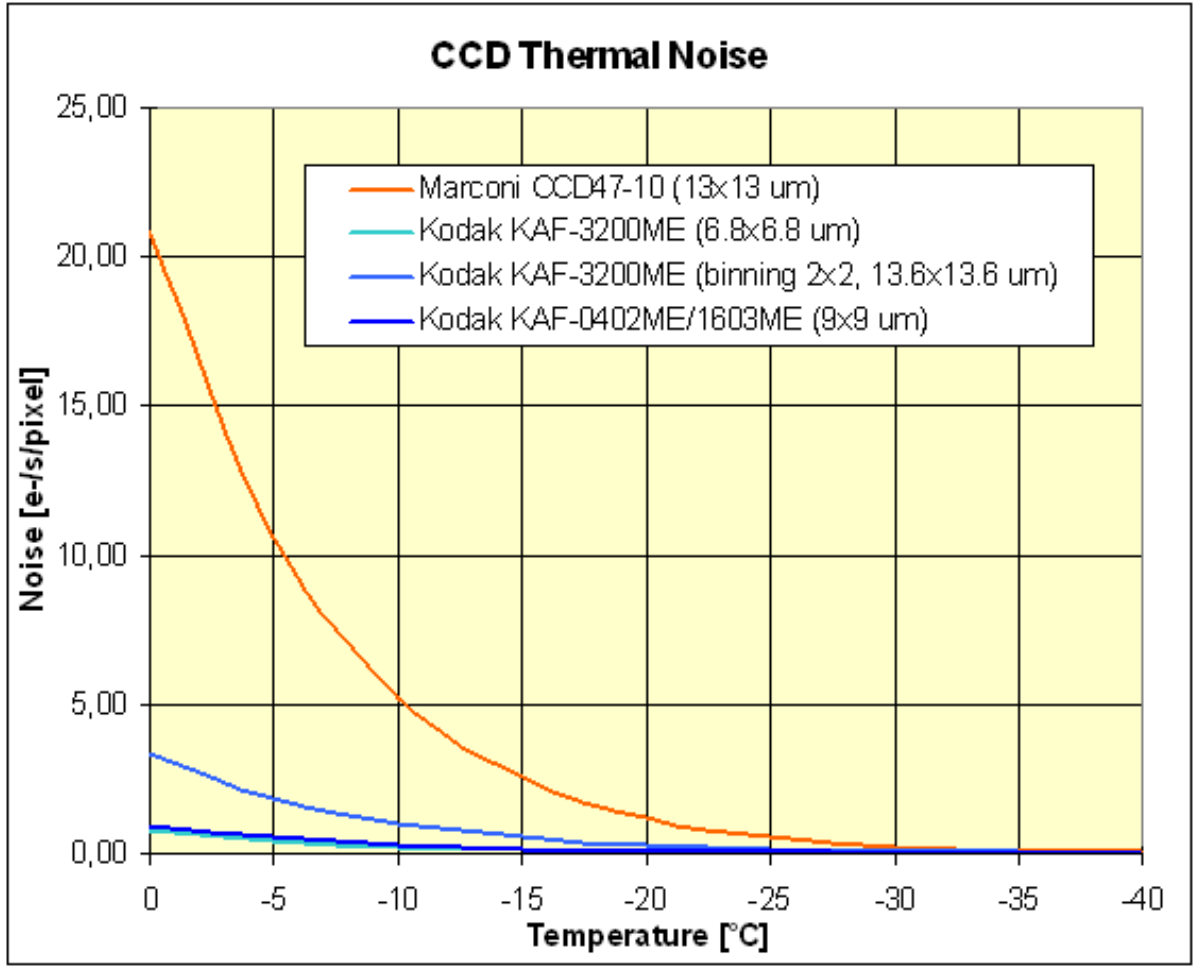}
      \caption{Thermal noise as a function of temperature for the sensor KAF-3200ME (pixel size $6.8\times6.8$ $\mu$m$^2$) 
      integrated in the CCD camera 
      used in this study (green solid line). The performances of other off-the-shelf devices are shown for comparison.}
         \label{darkc}
\end{figure}

For the purpose of the feasibility study we adopted existing instrumentation available 
at the OAVdA site. Most of the observations were carried out with a 250-mm Maksutov telescope 
(the optical scheme is shown in Fig.~\ref{optscheme}), with a focal lenght of 950 mm (f/3.8), 
on a german equatorial mount, and no autoguiding (while not common practice, this does not 
constitute a serious issue for the purpose of our study). 
The instrument is equipped with a front-illuminated Charged Coupled 
Device (CCD) camera Moravian G2-3200ME (sensor KAF 3200ME, area $2184\times1472$ pixel$^{2}$, 
pixel area 6.8x6.8$\mu$m$^{2}$), and a non-standard Astronomik $R$ filter (see Fig.~\ref{filters}). 
As shown in Fig.~\ref{qe}, the sensor reaches a 
quantum efficiency $QE\sim87$\% at this wavelenght. The CCD camera was operated with a nominal gain 
of 1 e$^{-}$/ADU (in $1\times1$ binning mode), with nominal read noise $N_r=10$ e$^{-}$ pixel$^{-1}$ (rms), 
dark current $N_D = 0.1$ e$^{-}$ pixel$^{-1}$ s$^{-1}$ at a temperature of $-40\,^{\circ}\mathrm{C}$  
and a full frame download time of $\sim$8.4 s through a USB 2.0 port. 
The CCD chip is cooled thermoelectrically by a two-stages system of Peltier modules which allows 
to reach a maximum temperature gradient $\Delta T\simeq50\pm0.1$\,$^{\circ}\mathrm{C}$ with respect 
to the ambient temperature. The chip temperature, recorded in the fits file headers, 
is typically stable to better than $0.5\,^{\circ}\mathrm{C}$. 
During the period of the characterization study (late spring/summer), 
the chip was cooled to temperatures of $-30\,^{\circ}\mathrm{C}$ or $-35\,^{\circ}\mathrm{C}$ 
(while the working temperature is usually set to $-40\,^{\circ}\mathrm{C}$ during winter time). 
As shown in Figure~\ref{darkc}, the typical values of thermal noise for our CCD camera 
remain well below 1 e$^{-}$ pixel$^{-1}$ s$^{-1}$ as far as the cooling temperature is set to 
$\lesssim -25\,^{\circ}\mathrm{C}$. 
The telescope-CCD configuration has a Field of View (FoV) of $52.10\times35.11$ arcmin$^{2}$ 
and a plate scale of $1.43^{\prime\prime}$/pixel. 

The OAVdA site also routinely utilizes two Ritchey-Chr\'{e}tien reflector telescopes, 
one 400 mm (f/7.64) and one 810 mm (f/7.9), the latter being the primary scientific instrument of the 
observatory. Both the instruments have equatorial open fork mounts. 

The 400 mm telescope will be used full-time together with the 250 mm during the next steps following 
this study, as detailed in \S~5. It is coupled to a front-illuminated CCD camera Finger Lakes 
Instrumentation Pro Line PL1001E with sensor KAF1001E and standard $BVRI$ filters. The camera sensor 
area is $1024\times1024$ pixel$^{2}$ with pixel area $24\times24$ $\mu$m$^{2}$. The final configuration is 
characterized by a FoV of $26.4\times26.4$ arcmin$^{2}$ with a plate scale of $1.55^{\prime\prime}$/pixel. 
The primary telescope is coupled to a back-illuminated CCD camera Finger Lakes Instrumentation 
Pro Line PL 3041-BB with sensor Fairchild Imaging 3041 and standard $BVRI$ filters. The camera 
sensor area is $2048\times2048$ pixel$^{2}$, with pixel area $15\times15$ $\mu$m$^{2}$. 
The system has a FoV of $16.3\times16.3$ arcmin$^{2}$ with a plate scale of $0.48^{\prime\prime}$/pixel. 
The detector is used routinely in binning mode $2\times2$, a choice driven by the need to 
avoid image oversampling, considering the typical seeing conditions at our site (see \S~\ref{seemeas}). 
While usually devoted to other projects, it will be used in the future to perform 
multi-band photometric observations within the context of the necessary follow-up scheme 
of potentially interesting transit candidates. 
Each observing station is equipped with a pc which is synchronized every 15 minutes via Internet 
to the Network Time Protocol (NTP) server of the Italian National Institute of Metrological Research. 
In the header of each frame, time is saved as Julian Date (JD) with six significant digits.
Table~\ref{table:1} summarizes the specifications of each of the three telescope systems.

\begin{figure}[h]
   \centering
   \includegraphics[scale=.35]{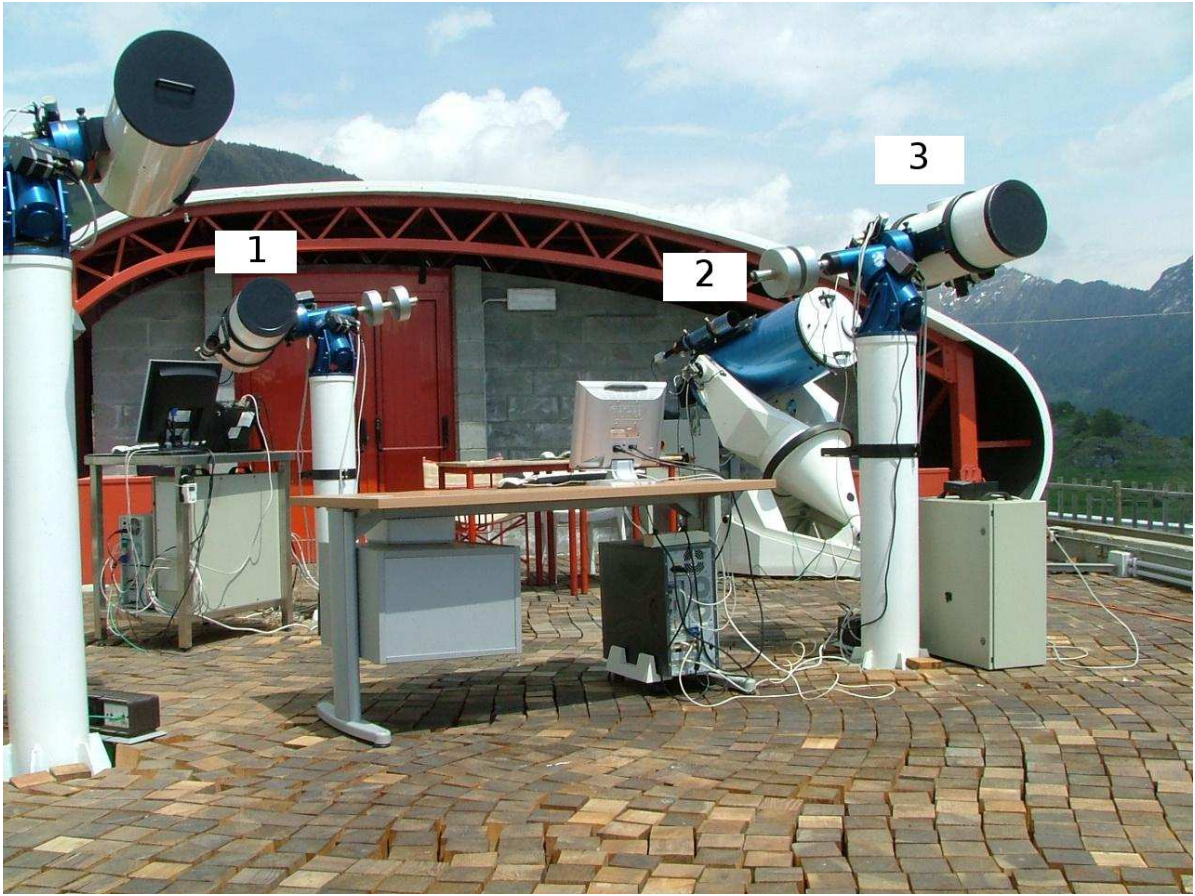}
      \caption{The OAVdA scientific platform from which the site characterization study 
      observations were carried out, and where an array of telescopes will be installed for a future 
      long-term photometric survey. Numbers 1 and 3 identify the two 250 mm Maksutov telescopes 
      used, respectively, to monitor the seeing with the DIMM technique and to acquire photometric 
      data of fields with stars hosting transiting extrasolar planets. In the middle (number 2) 
      is the 400 mm reflector telescope.}
         \label{terrace}
\end{figure}

\begin{figure}[h]
   \centering
   \includegraphics[scale=1.]{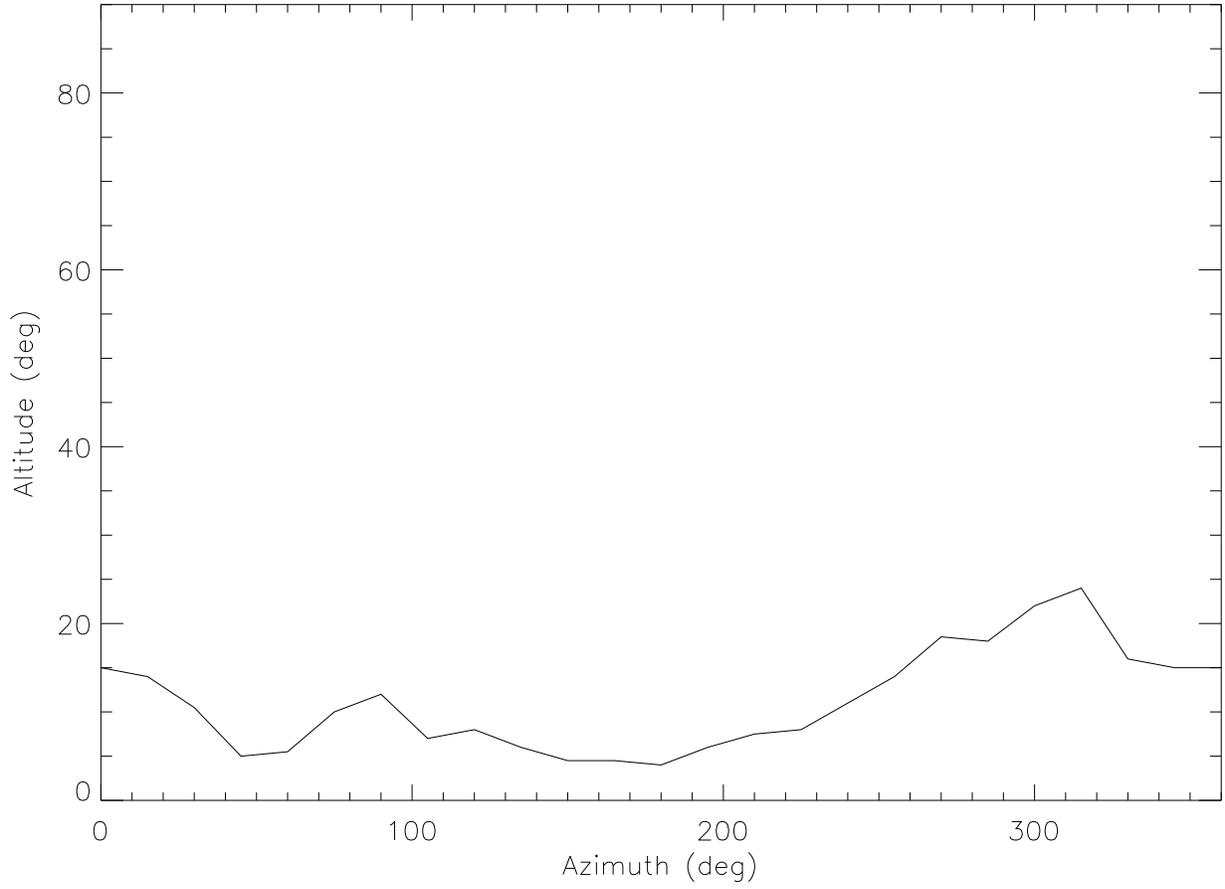}
      \caption{Minimum sky altitude useful for observations as a 
      function of the azimuth, before incurring into obstacles (observatory dome, 
      platform walls, and mountains). The profile was determined with a theodolite placed in the 
      center of the OAVdA platform which hosts the telescopes used in this study.}
         \label{horizon}
\end{figure}

The 250 and 400 mm telescopes are located in one of the platforms of the 
observatory. Fig.~\ref{terrace} shows a panoramic view of the telescopes location. 
We evaluated with a theodolite the minimum sky altitude above the horizon, as a 
function of the azimuth, that can be reached with a telescope placed in the middle of 
the platform before encountering any obstacle limiting the observations (mountains, 
platform walls, dome). The altitude-azimuth profile is shown in Fig.~\ref{horizon}. 
The most prominent obstacle is represented by the dome (with a 7-m diameter) hosting the 
primary telescope. The dome (not shown in Fig.~\ref{terrace}) is located next to the platform. 
It covers $\sim20-24$ deg in elevation, in the azimuth range $\sim$290-320 deg. 
This does not represent a serious constraint, however, as per standard procedure 
we observe targets until they reach a minimum altitude of 30 deg above the horizon.

\subsection{Telescope control system}\label{telescope} 

The 400 and 810 mm telescopes are managed by the control and data acquisition system 
CompactRIO which is based on National Instruments (NI) technology. The system is 
powered by a software with a graphical interface developed with the NI LabVIEW programming 
tools and it allows the remote control of the telescopes (the pointing is both automatic 
and in manual jog). The algorithms integrated in the control system were entirely 
developed by members of the OAVdA reasearch staff. The system is able to dialog, 
through the use of the LX200 protocol, with the most popular softwares used to control astronomical instruments.

\subsection{Seeing monitoring}\label{seemon}

\begin{figure}[h]
   \centering
   \includegraphics[scale=1.5]{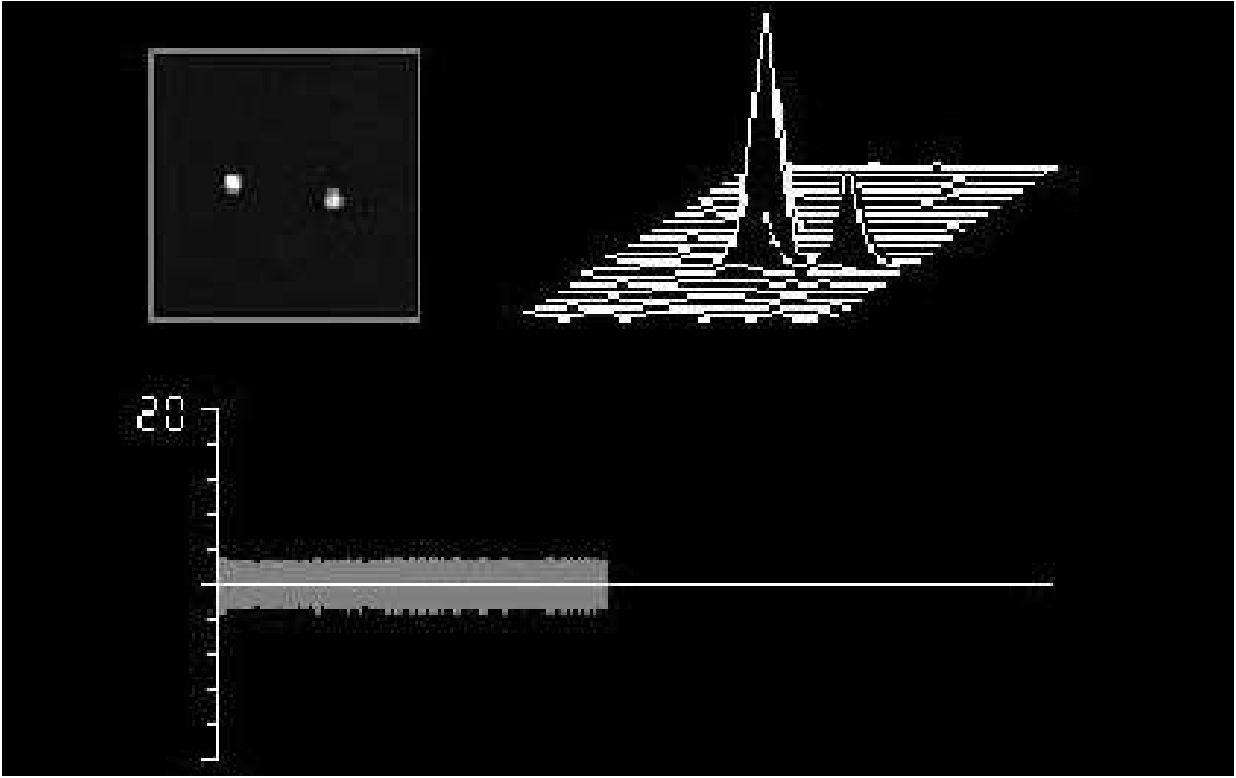}
      \caption{Freeze-frame from the SBIG-STV digital integrating video camera coupled 
      to a 250 mm Maksutov telescope to monitor the seeing with the DIMM method. 
      In the upper left corner the two images of the star used to estimate the seeing 
      are represented, as produced placing a two-hole Shack-Hartmann mask in front of 
      the telescope aperture.}
         \label{stv}
\end{figure}

During the period of the characterization study, each night of observation we monitored 
and recorded the seeing using the Differential Image Motion Monitor (DIMM) technique 
(Sarazin \& Roddier 1990), which minimizes the effects induced by the instrumentation 
and provides an accurate estimate of the level of the atmospheric turbulence. This technique 
requires the use of a two-hole mask (Shack-Hartmann mask) in front of the telescope aperture 
so that, slightly defocusing, two distinct and close profiles of one bright star are formed. 
Fig.~\ref{stv} shows a freeze-frame taken during a typical seeing monitoring session 
illustrating the DIMM procedure. A seeing estimate is then obtained measuring 
the relative motion of the two stellar profiles over time scales of ms. To monitor the 
seeing we used a second 250 mm f/3.8 Maksutov telescope (shown in Fig.~\ref{terrace}) 
equipped with a Shack-Hartmann mask and a SBIG STV digital integrating video camera which 
is able to apply in real-time the DIMM algorithm with high time sampling (typically, 
5 measurements per second). 
Data are automatically saved on a PC as ASCII files. No filter was used to perform these measurements 
(possibly resulting in an overestimate of the seeing).  

\section{Data processing and analysis}\label{dataproc}

\begin{figure}[h]
   \centering
   \includegraphics[scale=1.]{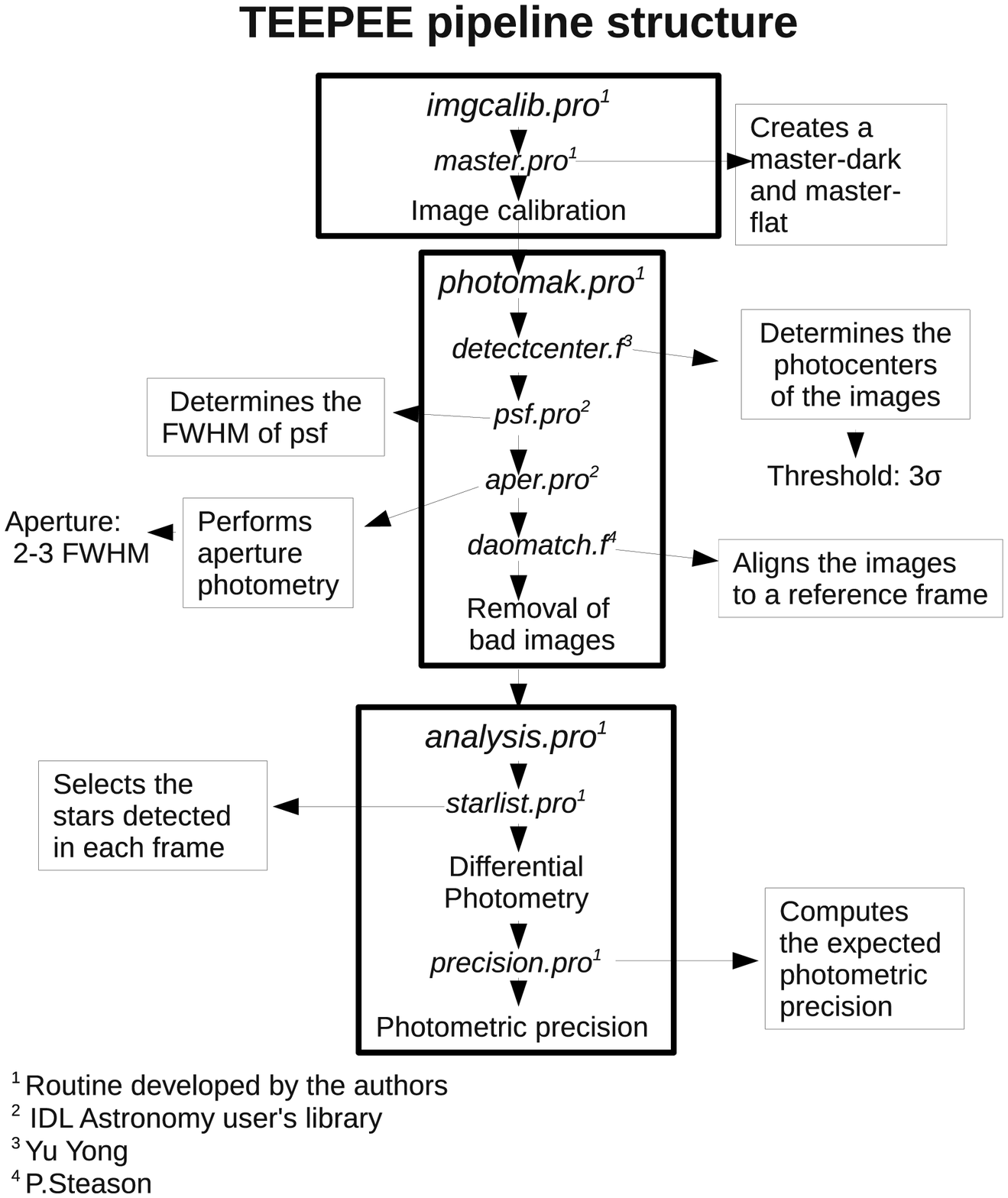}
      \caption{Flowchart summarizing the main tasks (and corresponding routines) performed 
      by the version 1.0 of TEEPEE, the pipeline for data reduction and processing developed 
      during the study presented in this paper. More details are given in the text.}
         \label{teepee}
\end{figure}

A fundamental element of the characterization study has been the development and testing of 
the automatic pipeline TEEPEE (Transiting ExoplanEts PipElinE) for the reduction and processing of the 
photometric data. TEEPEE is a software package written in IDL\footnote{IDL is a commercial programming 
language and environment by ITT Visual Information Solutions.~\url{http://www.ittvis.com/idl/}}, 
which utilizes publicly available software from the Astronomy Users' Library
\footnote{\url{http://idlastro.gsfc.nasa.gov/contents.html}} as well as external contributed FORTRAN 
routines. Two routines, {\tt aper} and {\tt getpsf}, are IDL versions of original IRAF routines
\footnote{IRAF is distributed by the National Optical Astronomy Observatory, which is operated by 
the Association of Universities for Research in Astronomy (AURA) under cooperative agreement 
with the National Science Foundation.}. The original portion of software is 
the one devoted to automatically perform ensemble differential aperture photometry on 
an user-specified stellar target. 
The various tasks performed by TEEPEE and the different routines used are 
summarized in the diagram shown in Fig.~\ref{teepee}. 
TEEPEE is organized in three main, sequential blocks, whose functionalities are described below.

\subsection{Image calibration}\label{calib} 

In the package {\tt imgcalib}, every raw FITS image acquired by the CCD (each 6.1 Mb in size) 
is initially dark-subtracted and flat-fielded. A master-dark image is created every 
month from the median values of each pixel of 70 dark images 
having the same exposure time of the science images\footnote{The number of dark images 
ensures minimization of the noise for the purpose of achieving a goal of a few milli-mag photometric 
precision. We are aware of the fact that taking dark images during each observing session might 
improve to some degree the quality of the calibration. The protocol will be changed in the future, 
by taking for example dark frames both at the beginning and at the end of the observing session}. 
The flats are prepared from 
the nightly averages of $\sim$15 high-S/N exposures at twilight, diffused by a white opal of polistyrene. 

\subsection{Photometric \& astrometric processing}\label{process}

The first task of the pipeline package {\tt photomak} is the search for the photocenters in each image. 
This is performed automatically by the subroutine {\tt detectcenter}, which selects each pixel 
with an ADU value greater than 3 times the rms of the sky background. The list of $x$ and $y$ 
coordinates of the photocenters are used as input for the subroutine {\tt psf}, which calculates 
the average $FWHM$ of the point-spread function (PSF) of each detected star in every image. 
Then, the {\tt aper} subroutine is applied to perform single-aperture photometry on each detected object, 
with the radius of the user-provided circular aperture typically set to $2-3$ times 
the average value of the $FWHM$. This routine counts the ADU values for each pixel in the circular aperture, 
while partial pixels are handled by numerical integration. Once the photocenter coordinates and the instrumental magnitudes 
are listed, the {\tt daomatch} subroutine alignes each image to a reference frame 
(usually, the first image of each series) by performing a triangulation. If in one image the stars 
found are less than 20$\%$ of those detected in the reference frame or the alignment 
procedure fails (mostly because of high variability in the sky transparency), 
then the image is automatically discarded. 

\subsection{Data analysis}\label{analysis}

The heart of TEEPEE, for the purpose of this study, is the package {\tt analysis}. 
It performs those operations which are necessary to eliminate, 
when possible, all the systematics which cause the degradation of the photometric 
precision, and outputs the photometric light curves for stars detected in a every image of a field.
At first, by using the subroutine {\tt starlist} only the stars that have been 
detected in every frame are selected and added to the final list. All objects that, 
due to telescope drifts exited the CCD FoV, or are too weak (i.e. below 3$\sigma$ of the sky background 
level) in one or more frames are discarded. The typical drift of the 250 mm telescope 
is nearly 10 pixel hr$^{-1}$ and 20 pixel hr$^{-1}$ in the X and Y direction, respectively.
Considering, for example, the fields of WASP-3 and HAT-P-7 which we monitored for this study 
(see \S~\ref{photper}), on average $\sim15\%$ of the stars appearing in the first frame of 
a series are lost because of drifts or veils, resulting in some 550-600 good stars detected on average 
in every frame during a night. Finally, the corrections due to atmospheric extinction are applied. 
The equation giving the measured magnitudes as a function of the magnitude above the atmosphere is:

\begin{equation}
\centering
m_I=m_o+k\times AM,
\label{eqk}
\end{equation}

where \textit{k} denotes the atmospheric extinction coefficient, \textit{m$_{I}$} the instrumental 
magnitude, \textit{m$_{o}$} the magnitude above the atmosphere and \textit{AM} the air-mass 
($AM=1$ when the object is located at the local zenith). The standard procedure requires 
that slope \textit{k} and intercept $m_o$ be evaluated every time from observations of photometric 
standard stars, such as Landolt stars. Given the purely differential approach to the 
photometric measurements in this study, we proceeded to calculate the extinction coefficient 
using all the \textit{n} reference stars chosen in the field of each target. This is done by solving a 
global system of \textit{n}$\times$\textit{$N_i$} linear equations of the form of Eq.~\ref{eqk} 
through the IDL routines {\tt SVDC} and {\tt SVSOL}, where \textit{$N_i$} is the number of good 
frames of the series. The solution is represented by a global value for 
\textit{k} and \textit{$N_i$} values for \textit{$m_0$}. As comparison stars we selected the \textit{n} 
brightest stars detected in every frame with $m_I\leq13$. For the purpose of this study, the $R$-band 
measurements of $k$ are used to gauge the level of sky transparency (see \S~\ref{nightsky}). No dedicated 
device (e.g., CAVEX, S\'anchez et al. 2007) for the monitoring of extinction was used.

The rms of each new light curve obtained after the air-mass correction provides quick information 
about the stability of the atmospheric transparency during the night, but more accurate light 
curves are provided by the ensemble differential photometry.
For each frame \textit{i}, we used as the reference magnitude \textit{m$_{rif}^{i}$} the average 
magnitude of the \textit{n} reference stars

\begin{equation}
\centering
m_{rif}^{i}=\frac{\sum_{k=0}^{n} m_{k}^{i}}{n}.
\end{equation}

{m$_{rif}$} is then subtracted to the magnitude of the user-defined target 
\textit{M$_{target}$}, obtaining the difference $\Delta m^{i}=m_{rif}^{i} - m_{target}^{i}$. 
The procedure is iteratively repeated for all the reference stars too, 
using as new references the remaining $n-1$ stars. We thus obtain a series of 
differential light-curves for all the stars selected, which can be inspected for 
variability. The few bright variables identified with this approach (see \S~\ref{newvars}) 
are not discarded at this stage, as their impact on the photometric precision is 
minimized by the selection of dozens of comparison stars. 

As discussed, the version 1.0 of the TEEPEE pipeline can perform automatically 
several basic tasks. These have been tested with promising results, as described in the following 
sections. However, the pipeline is still far from being completed. Further refinements will 
guarantee improved functionality and a higher degree of robustness. In particular, a photometric 
survey finalized to the detection of planetary transits must take into account the effects 
of 'red noise', which can strongly affect the sensitivity thresholds~\citep{Dam-Pont06}. 
This feature will be included in future upgrades of the software package in preparation of the 
official start of the photometric survey in the near future. 
The pipeline will grow further to implement algorithms for the reduction of the noise and 
systematic variations in the light curves (detrending), 
such as the Trend Filtering Algorithm (TFA,~Kov\'{a}cs et al. 2005) or SysRem 
(Tamuz et al. 2005), as well as algorithms for the detection of periodic transit-like signals, 
such as the widely used Box-fitting Least Squares (BLS) method~\citep{Dam-Kova02}. 

\section{Site characterization study: results}\label{results}

The main OAVdA site characterization campaign was carried out during the period May-August 2009. 
As in any search for adequate sites for new observatories, we focused on the determination 
of observing conditions such as night-sky brightness, number of useful nights, photometric 
precision, and seeing. 

\subsection{Night-sky brightness and transparency}\label{nightsky}

We have utilized measurements of the zenithal night-sky brightness collected between 
May 2007 and July 2007 during seven moonless nights in collaboration with the italian 
Agency for Environmental Protection (ARPA), as part of an extended monitoring program of 
the light pollution in the Aosta Valley. The data were obtained using the 400 mm 
reflector telescope and a CCD camera equipped with a $V$-band filter. All measurements 
were calibrated using standard reference stars. The $V$-band night-sky 
brightness ranged between 21.12 and 21.61 mag arcsec$^{-2}$, with a mean and standard 
deviation of $21.29\pm0.16$ mag arcsec$^{-2}$. These numbers are typical 
of those of very good, very dark observing sites (e.g., see Table 4 in~\cite{Dam-Mol09}). 
During the same period, the $V$-band extinction coefficient ranged from 0.116 mag to 0.420 mag, 
with a mean and standard deviation of $0.26\pm0.13$ mag. This value compares reasonably well with 
the $V$-band $k$ estimates during summer available in the literature for other sites (e.g., S\'anchez et al. 2007, 
and references therein). The mean and standard deviation of our own $R$-band $k$ value derived 
within the context of the site characterization campaign are $0.165\pm0.090$ mag. They are in good 
agreement with the analytical estimate, 0.146 mag, from the extinction curve formula in S\'anchez et al. (2007) 
based on the $V$-band value of $k$ reported above. The typical conditions of sky transparency are thus in line 
with what to expect from a good site, although the spread around the mean conditions is 
relatively large. Given the limited time span of the observations, we cannot comment on 
seasonal variations of the extinction at the OAVdA site, nor are we in a position to investigate 
in detail the relative role of various contributors (e.g., dust and other aerosol particles, 
ozone).

\subsection{Fraction of useful nights}\label{nights}

In order to estimate the number of useful nights/yr, we relied upon on a 3-yr baseline of 
data (from 2006 to 2008) collected by a weather station located next 
to the Observatory, while for the year 2009 we also used our own internal logs. Over 
this time span, fully clear nights conditions are statistically realized 142 night/yr 
(39$\%$), while an additional 67 nights/year (18$\%$) can still be considered useful, 
assuming a typical cloud cover of $<50$\% of the useful observing time (conservative upper limit). 
As for the year 2009, during which we carried out the feasibility study, we registered 
similar conditions: uninterrupted observations could be carried out 39.5$\%$ of time, 
while for an additional 22$\%$ of the nights the cloud cover did not exceed 50$\%$ of the 
useful observing time. These figures compare reasonably well with other observatory 
sites in Europe.

\subsection{Photometric performance}\label{photper}

\begin{figure}[h]
   \centering
   \includegraphics[scale=.8]{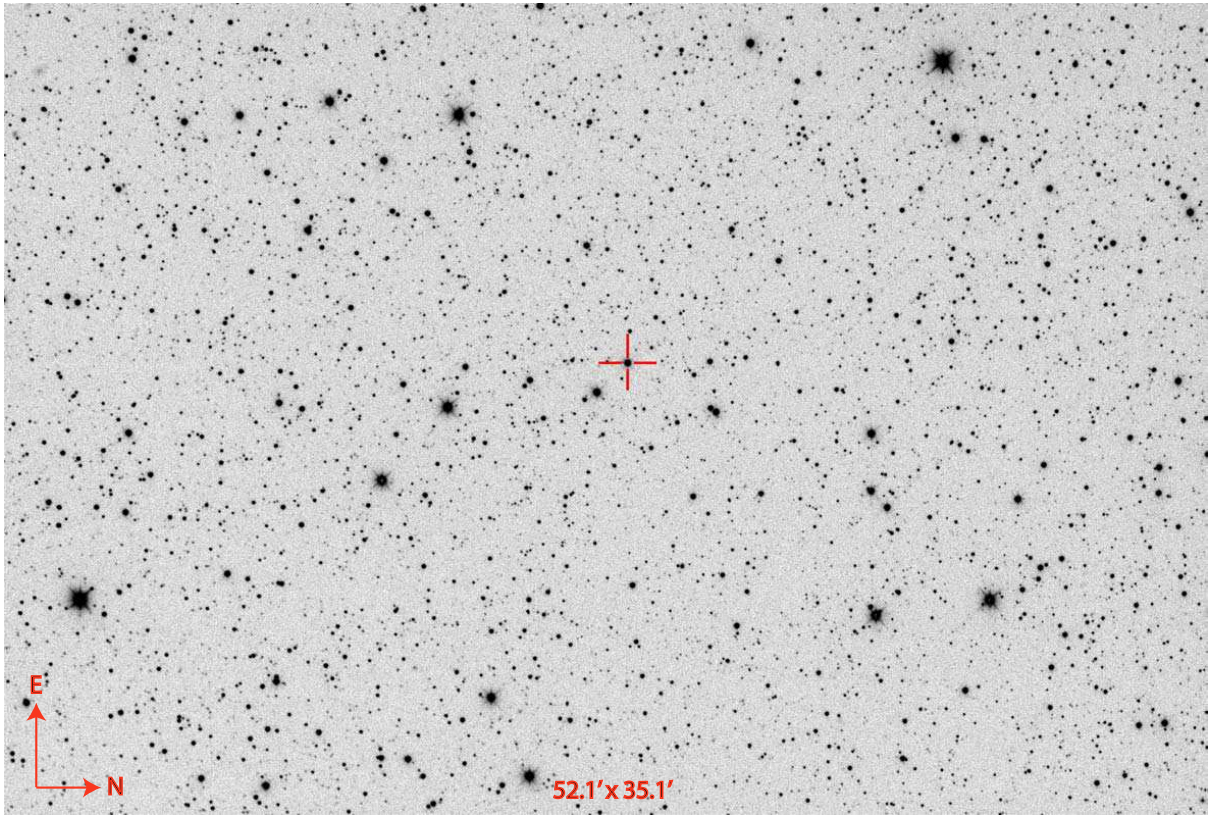}
      \caption{A typical FITS image obtained with the 250 mm telescope. 
      In this example the field of the star HAT-P-7 is shown, indicated 
      with a cross (Exposure time: 60 s).}
         \label{field}
\end{figure}

In order to estimate the achievable photometric precision at the OAVdA site, given 
the above mentioned instrumental setup, the centerpiece of the site characterization 
study consisted in the monitoring of three stellar fields centered on the transiting planet 
hosts Gliese 436~\citep{Dam-But04}, WASP-3~\citep{Dam-Pol07}, and HAT-P-7~\citep{Dam-Pal08}. 
The targets were observed during both in- and out-of-transit phases during the period 
May-August 2009.  All observations were 
performed with the CCD set up in $1\times1$ binning mode and at the focus of the telescope, and the 
exposure times were chosen at the beginning of each session (without being subsequently modified) 
in order to guarantee an optimum signal-to-noise ratio ($S/N$) for the target, while avoiding 
saturation. Typical exposure times were 22 s for Gliese 436 ($V$=10.68 mag), 
the brightest target at $R$ band, and ranged between 35 and 60 s for both WASP-3 ($V$=10.48 mag) 
and HAT-P-7 ($V$=10.46 mag), given variable conditions of air mass and sky transparency. 
Fig.~\ref{field} shows an example of the quality of the 
images we collected. The whole database for this characterization study comprises 
$\sim$10,300 good images,  corresponding to $\sim6$3 GB of data.

\begin{figure}[h]
   \centering
   \includegraphics[scale=1.]{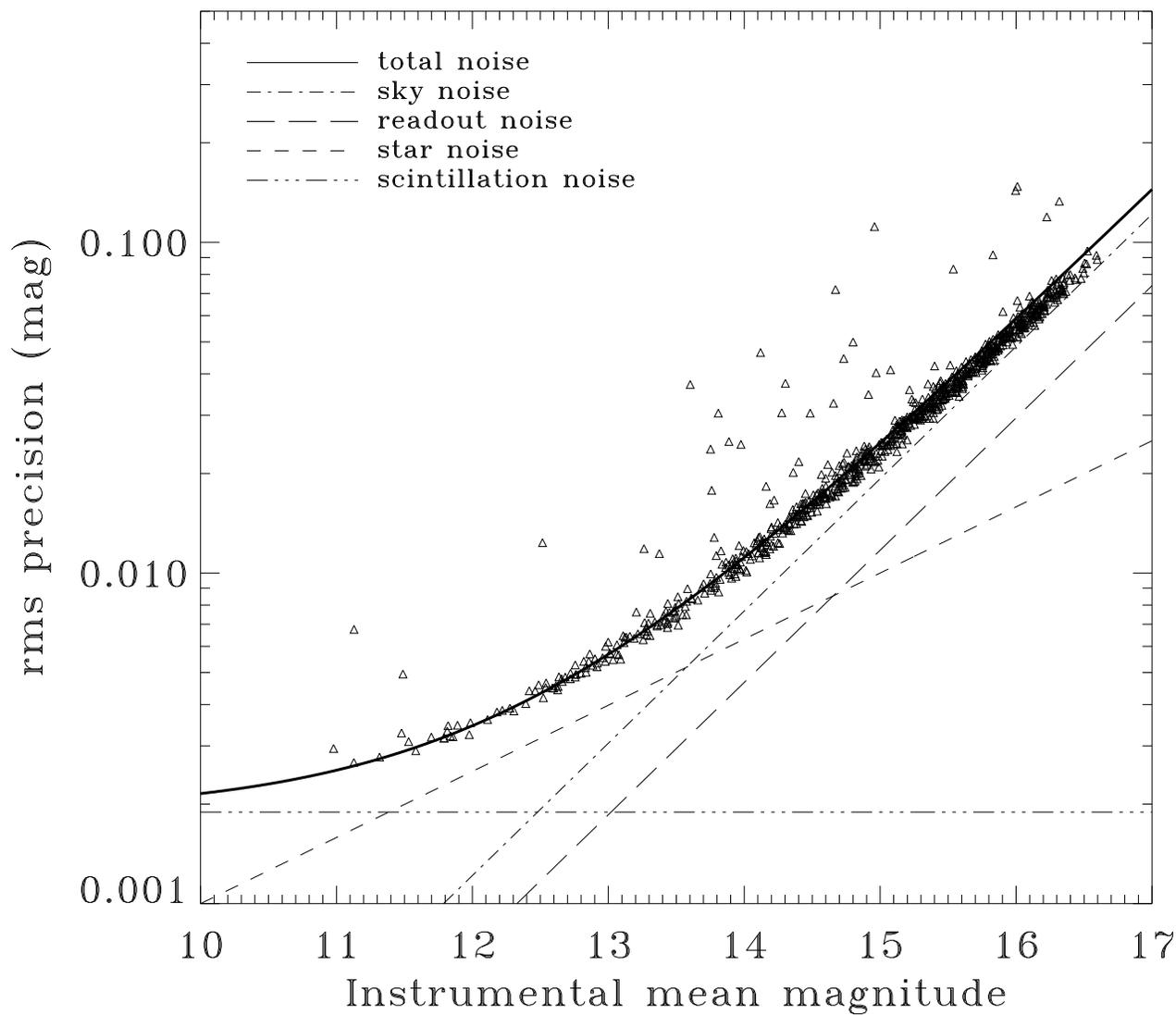}
      \caption{Photometric errors (rms) vs. instrumental mean magnitude in the field of the planet-hosting star WASP-3 
      during a good observing night (and for an exposure time of 35 s). 
      The various contributions to the expected photometric noise, according to Eq.~\ref{precphot}, 
      are also shown.}
         \label{rmsvsmag}
\end{figure}

We show in Fig.~\ref{rmsvsmag} the scatter in our differential photometry as a 
function of apparent magnitude in the field of the star WASP-3, 
for a night representative of the typical performance 
achieved. The results are compared to a theoretical noise estimate~\citep{Dam-NUCH08} 
which includes contributions from the object (Poisson noise), sky, readout 
noise, and scintillation:

\begin{equation}
\centering
\sigma_{ph} = \frac{\sqrt{N_\star +\sigma ^2_{scint}+n_{pix}\times\left(N_S+N_D+N_R^2\right)}}{N_\star}
\label{precphot}
\end{equation}

In Eq.~\ref{precphot}, $ N_\star $ is the number of received photons from the star 
during the exposure, $ n_{pix} $ is the number of pixels in the 
photometric aperture, $ N_S $ is the average number of photons per pixel of the 
sky background (we used a number typical of the average sky conditions registered over the duration 
of our campaign), $ N_D $ is the dark current contribution in e$^{-}$ pixel$^{-1}$ 
(from \S~\ref{hardware}), $ N_R $ is the rms of the CCD read noise in e$^{-}$ pixel$^{-1}$ 
(also from \S~\ref{hardware}), and $\sigma_{scint}$ is the noise term for the 
atmospheric scintillation. The expression used for $ \sigma_{scint}$ is (Dravins et al. 1998):

\begin{equation}
\centering
\frac{\sigma_{scint}}{N_\star} = 0.09\frac{AM^{3/2}}{D^{2/3}\sqrt{2t}}\exp\left(-\frac{h}{8}\right),
\label{scint}
\end{equation}

where $ AM $ is the air-mass (we adopted a mean value of 1.5), $ D $ is the telescope 
aperture in cm, $ t $ is the exposure time in s, and $ h $ is the height in km of the 
observing site above the sea level. 

For bright stars, the dominant sources of error are the scintillation and shot 
(Poisson) noise from the target star itself. Notice the excellent agreement with 
the expected scatter values for magnitudes $R\lesssim13$ mag. Among the most 
likely sources of additional scatter in the observations with respect to the 
theoretical predictions we list guiding errors and possible contributions from 
centering errors during aperture photometry. For the 
brightest objects we achieve a nightly rms precision of $\sim0.003$ mag 
(during some nights the recorded scatter for these stars was even lower). 
For fainter stars, sky photon noise is the dominant source of scatter. 
Some outliers can be seen in the dataset. One of them corresponds 
to a newly discovered variable (probable $\delta$ Scuti-type) star \citep{DAM10}. 

\begin{figure}[h]
   \centering
   \includegraphics[scale=1.]{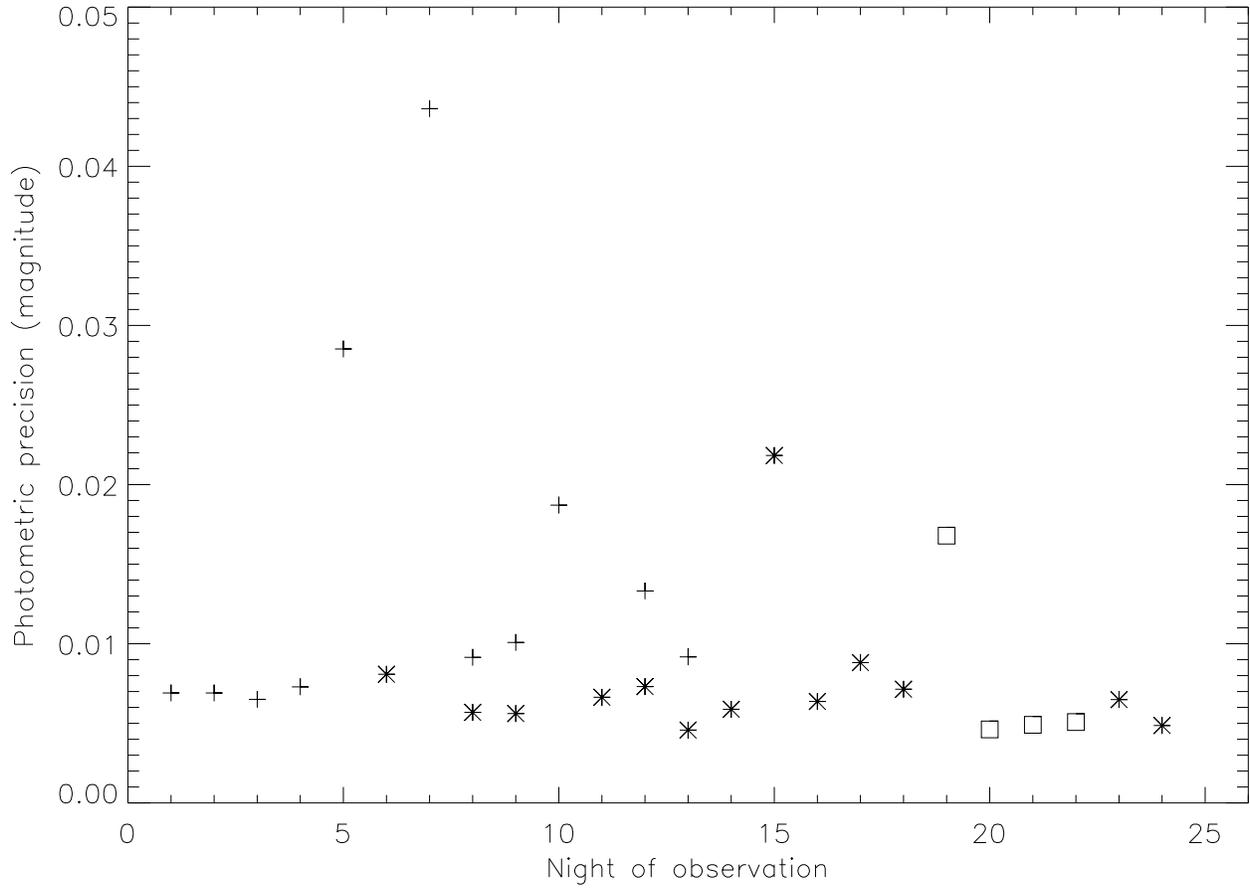}
      \caption{The nightly-averaged photometric precision, 
      calculated for stars with magnitudes $<13$ mag, is plotted for 
      each night of observations. Different symbols are used to distinguish 
      the three stellar fields monitored during the characterization study: 
      a plus denotes the field of the star Gliese 436, an asterisk is used 
      for WASP-3, while squares correspond to HAT-P-7.}
         \label{precision}
\end{figure}

To evaluate the photometric precision of each night of observation we adopted the mean rms calculated for stars 
with magnitudes $R\lesssim13$ mag. In Fig.~\ref{precision} we show how the nightly-averaged 
rms varied during the period of observation. Different symbols are used to distinguish 
the various star fields that were monitored, as described in the figure caption. 
The median rms over the whole duration of the characterization study is 0.006 mag 
for magnitudes brighter than 13 mag. The results shown in Fig.~\ref{rmsvsmag} can 
then be considered representative of a typical night at our site. 
Equally important is the fact that the night-to-night 
variations are $\sim0.002$ mag (disregarding the few nights with large, $> 0.01$ mag 
scatter due to poor weather conditions). These results demonstrate that it is 
possible to achieve competitive precision at the OAVdA site over an extended 
period of time, a particularly encouraging result given the still non-optimized hardware 
system used during this study.

\subsection{Seeing measurements}\label{seemeas}

As described in \S~\ref{seemon}, seeing monitoring has been carried out in conjunction with the 
photometric measurements over the whole duration period of the site characterization study. 
For the purpose of this work, its role has been quantified in terms of its median 
value over the timespan of the measurements, its typical stability during a night, and 
its correlation with the photometric performance achieved. 
We note that, given the limited time duration of the measurements, we cannot comment 
in a statistically significant fashion on possible seasonal seeing variations, or in detail 
on its possible dependence on other meteorological parameters, such as wind speed and 
direction. 

\begin{figure}[h]
   \centering
   \includegraphics[scale=1.]{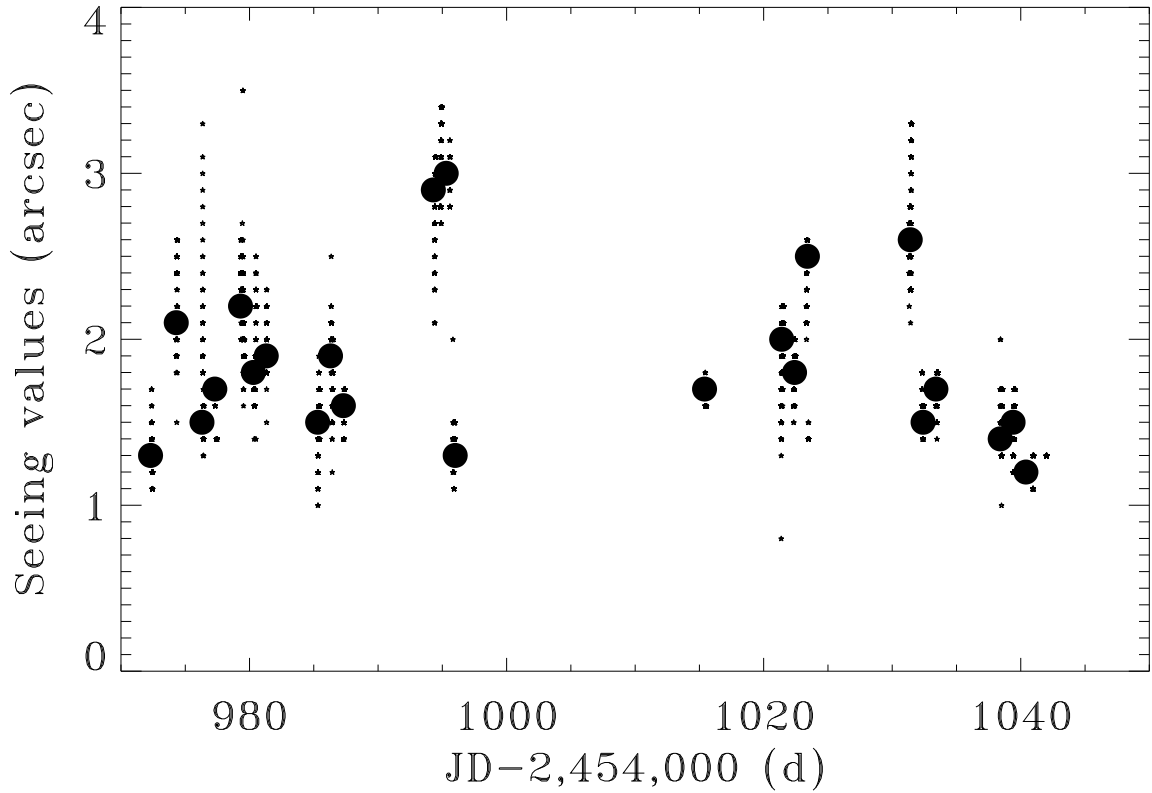}
      \caption{The distribution of seeing measurements over the 
      whole period of observation. The small black stars show 7864 individual data points 
      (many of them overlapping due to the seeing values having been approximated to the 
      first significant digit), each 
      corresponding to an average seeing value over a 1 min interval, rounded to 0.1$^{\prime\prime}$. 
      The large filled circles indicate the median seeing value for each night.}
         \label{seeing}
\end{figure}

Fig.~\ref{seeing} shows the distribution of seeing values as a function 
of the night of observation. The median seeing turns out to be  
$\sim1.7^{\prime\prime}$ over the whole period. In most nights, the seeing appeared 
rather stable, typically remaining within 20\% or so of the reference median value. 

\begin{figure}[h]
   \centering
   \includegraphics[scale=1.]{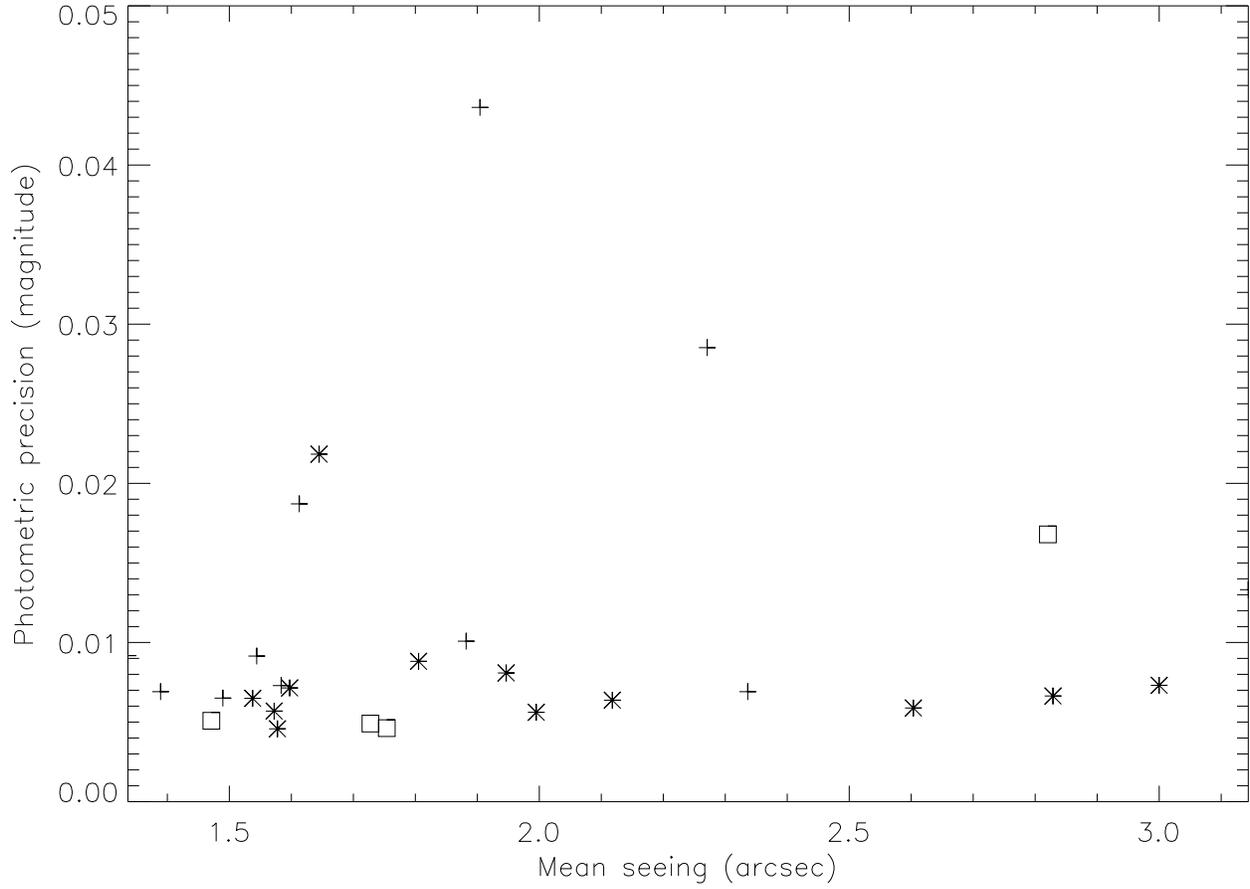}
      \caption{The comparison of the nightly-averaged photometric precision 
      to the corresponding mean seeing does not reveal an evident correlation 
      between these two variables for $FWHM$ values $\leq3^{\prime\prime}$, given our intrumental set up. 
      Symbols are as in Fig.~\ref{precision}.}
         \label{precsee}
\end{figure}

Indeed, the overall result is that the OAVdA site does not appear competitive 
with respect to other observing sites in the world (see, e.g., Table 5 in Moles et al. 2009). 
However, as illustrated in Fig.~\ref{precsee} the photometric precision is essentially seeing 
independent, for values up to $3^{\prime\prime}$. On the basis of this result we can conclude 
that the typical seeing measured at the OAVdA site does not represent a limiting 
factor for the precision of the stellar differential photometry (with the instrumental 
set up we utilized), for magnitudes $R\lesssim13$ mag. This is in accordance with the promising 
results found applying the telescope defocusing technique for high-precision 
photometry of transiting extrasolar planets (i.e. Southworth et al., 2009a, 2009b, 2009c), 
which can be carried out for the brightest stars even with small-size telescopes. 

\subsection{Sample light curves of known transiting systems}\label{samplelc}

\begin{figure}[h]
   \centering
   \includegraphics[scale=.6]{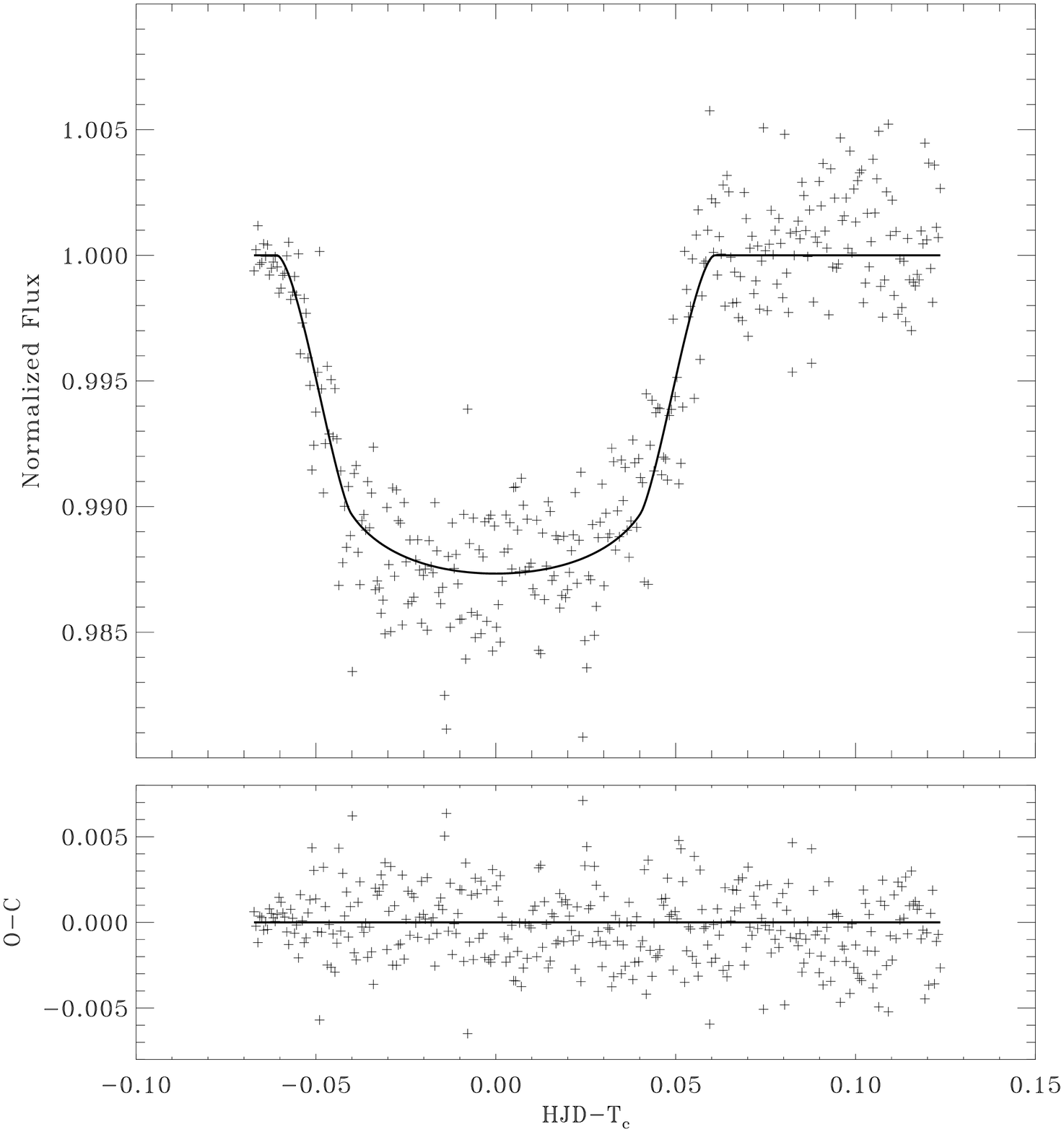}
      \caption{Top: transit light-curve of WASP-3b during the night of UT 28 July 2009. 
      The time of transit center occurred at T$_C=2455041.411$ HJD.  
      The 358 35-sec exposures were collected over 4.6 hrs. The best-fit curve is superposed. 
      Bottom:  the post-fit residuals exhibit an rms of $0.0021$ mag}
         \label{wasp3}
\end{figure}

\begin{figure}[h]
   \centering
   \includegraphics[scale=.45]{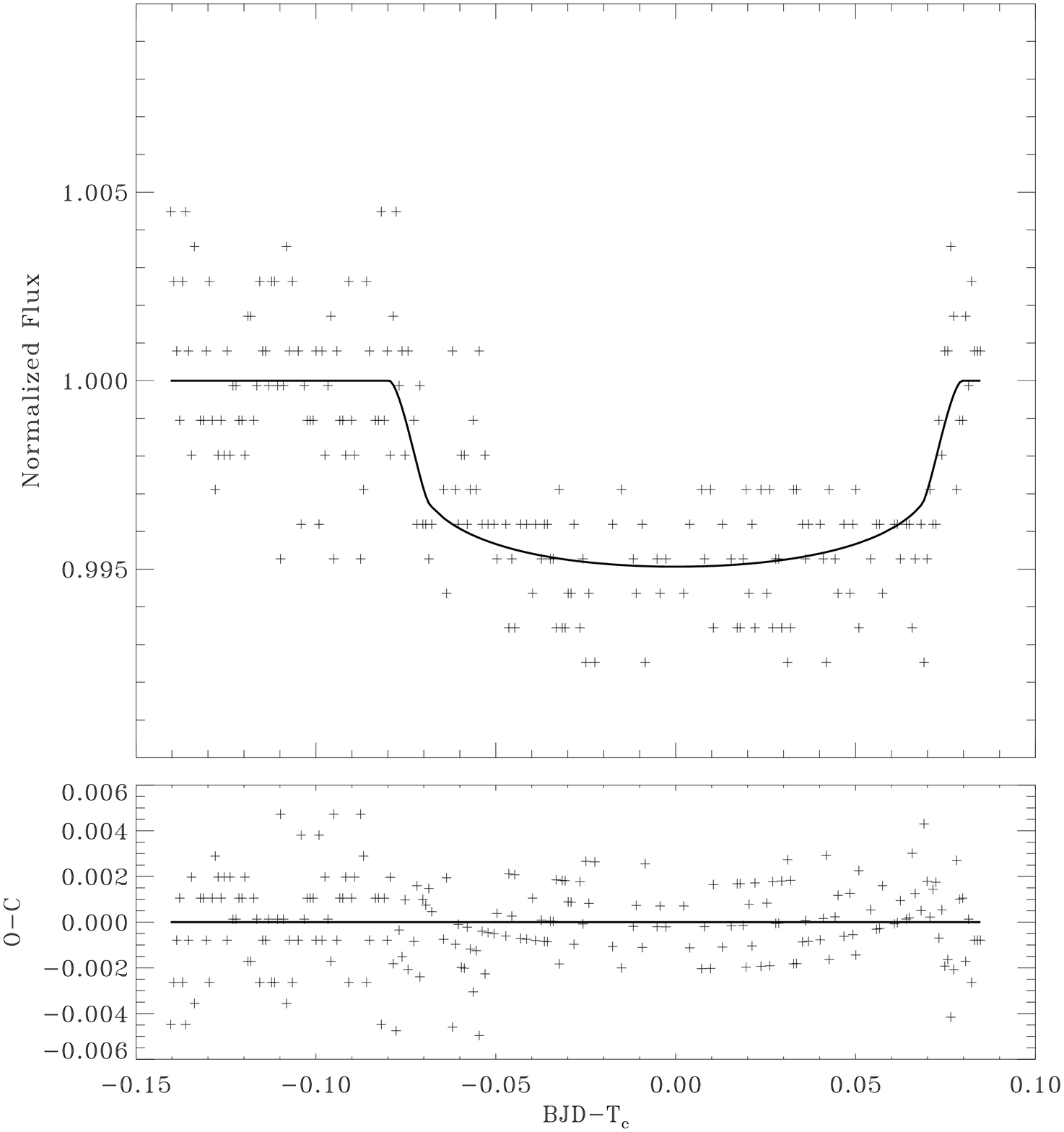}
      \caption{Normalized light-curve of HAT-P-7 showing the shallow 
      transit of its Jupiter-mass companion HAT-P-7b (UT 18 July 2009, 260 60-sec 
      exposures collected in 4.6 hrs). The time of transit center occurred at T$_C=2455031.5222$ BJD. 
      The best-fit curve is superposed. 
      Bottom:  the post-fit residuals exhibit an rms of $0.0018$ mag. }
         \label{hatp7}
\end{figure}

As examples of the quality of the photometric data achieved with the present 
instrumental setup, we present two light curves showing the transits of the extrasolar 
planets WASP-3b (Fig.~\ref{wasp3}) and HAT-P-7b (Fig.~\ref{hatp7}). As for Gliese 436b 
(whose data are not shown here), unfavourable wheater conditions prevented us from observing 
a full transit during the timespan of the characterization study. 

WASP-3 ($M_\star = 1.24\,M_\odot$, $R_\star=1.31\,R_\odot$), which hosts an $M_p=1.76\,\,M_J$, 
$R_p=1.31\,R_J$ planet with a period $P=1.846834$ days, was monitored for 19 nights. The rather good phase 
coverage allowed us to observe four transits of WASP-3b. Fig.~\ref{wasp3} 
shows the full transit observed during the night of UT 28 July 2009. 
Superposed to the normalized flux is the best-fit curve, obtained as solution of 
a non-linear least square problem using the Levemberg-Marquardt algorithm and a merit function 
based on the formalism of \citet{Dam-Mand02} with adjustable quantities the inclination angle $i$, 
the ratio of radii $r=R_p/R_\star$, and the impact parameter $b=(a/R_\star)\cos i$ (where $a/R_\star$ is the 
orbital semi-major axis expressed in units of the stellar radius). 
We assumed a quadratic limb darkening law with $R$-band coefficients $u_1$=0.2357 and $u_2$=0.3779 
(Claret 2000), appropriate for the stellar parameters in Pollacco et al. (2008). 
The sub-panel of Fig.~\ref{wasp3} shows the post-fit residuals, that exhibit an rms of 0.002 mag. 
The relatively large transit depth (0.013 mag) allowed us to derive formally precise 
values for the system parameters. The best-fit values and their associated $1-\sigma$ uncertainties 
derived from the covariance matrix of the solution are as follows: $b=0.665\pm0.003$, $i=81.24\pm0.06$ deg, and 
$r=0.1091\pm0.0006$. There appears to be some discrepancy (a less central transit) with respect to the values 
reported in the discovery paper of Pollacco et al. (2008) and in the work of Gibson et al. (2008). 
However, the results of this exercise are shown here only for illustrative purposes. 
Our results are undoubtedly affected by the non-optimal choice of the limb-darkening parameters 
(we recall observations were taken with a non-standard $R$ filter), by the lack of a proper 
treatment of red noise, and by the almost complete absence of data gathered before ingress.  

The star HAT-P-7 ($M_\star = 1.47\,M_\odot$, $R_\star = 1.84\,R_\odot$) was selected as a challenging 
target for our test because its transiting planet ($M_p=1.8\,M_J$, $R_p=1.42\,R_J$, 
$P=2.2047298$ days) produces a photometric transit characterized by a shallow depth ($\sim$0.007 mag), 
which is close to the detection limit for our present instrumental and data reduction and 
analysis setup. HAT-P-7 is located in the field of view of the Kepler space telescope, that monitored the 
star for ten days during its commissioning phase, deriving an exquisite-quality 
optical phase curve~\citep{Dam-BOR09}. In Fig.~\ref{hatp7} we show the transit observed 
at OAVdA the night of UT 18 July 2009. We could not observe a complete 
transit (the planet's egress is barely visible due to the approaching sunrise which forced 
us to interrupt the exposures),  and during the night the sky transparency varied significant due 
to the presence of veils. However, the transit was clearly detected, and by using an 
aggressive light-curve analysis, including iterative $\sigma$ clipping for the removal 
of outliers, we managed to 'clean' the dataset so as to perform an analogous exercise to that 
carried out in the case of WASP-3. The light-curve fit, assuming a quadratic limb darkening law 
with $R$-band coefficients $u_1$=0.3980 and $u_2$=0.2071, appropriate for the stellar parameters in 
P\'al et al. (2008), resulted in the following solution: $b=0.321\pm0.006$, $i=85.93\pm0.08$ deg, and 
$r=0.064\pm0.002$. In this case, the main difference is that of a value of $r$ which is about 15\% 
smaller than that reported in the literature. This is due to the shallower transit depth 
($\sim5$ mmag) recovered after we applied the procedure to remove outliers. Again, we stress the illustrative 
purpose of the orbital fit carried out, the main aim here being simply that of showing that the transit 
could clearly be detected. This represents a promising result in the perspective of our 
long-term photometric monitoring program to be started in the near future with 
upgraded instrumentation and an improved data reduction and analysis pipeline.

\subsection{New variable stars}\label{newvars}

\begin{figure}[h]
   \centering
   \includegraphics[scale=1.2]{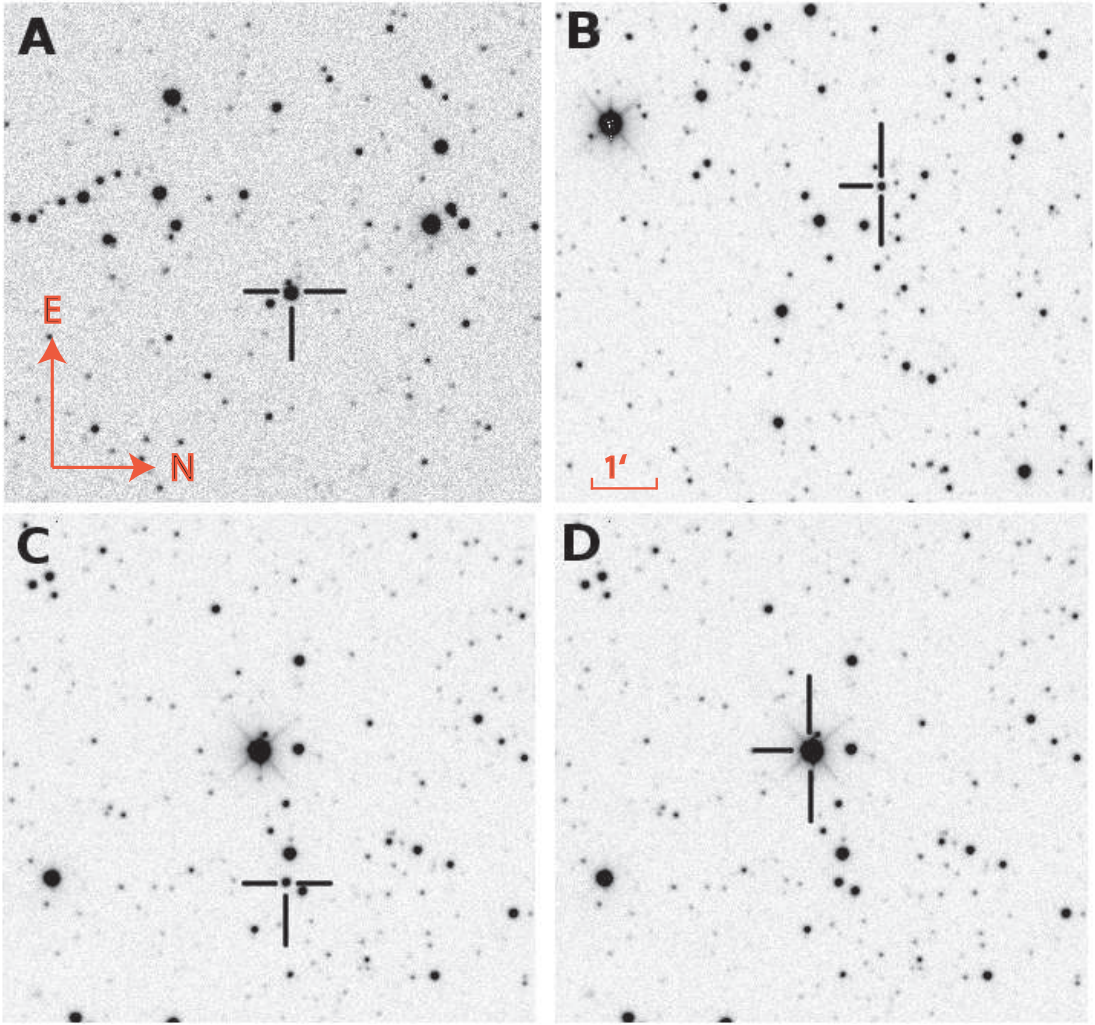}
      \caption{The new variable stars discovered at OAVdA and presented in this work. 
      \textit{A}: GSC2.3 N208000215. \textit{B}: GSC2.3 N2JH035417. 
      \textit{C}: GSC2.3 N2JH000428. \textit{D}: GSC2.3 N2JH066192.} 
         \label{variables}
\end{figure}

A natural byproduct of high-precision, high-cadence photometry 
collected for hundreds of stars over a timespan of several weeks, 
as is the case for the WASP-3 and HAT-P-7 stellar fields monitored 
in the context of our site characterization study, is the discovery 
of variable sources. A handful of previously unknown variable stars 
with relatively large amplitude variations ($\Delta V \gtrsim 0.05$) 
were uncovered during this study. Two of them are presented and discussed 
in Damasso et al. (2010). Here we report the detection of four new 
binary/multiple-star systems showing eclipses. The fields centered on the 
variables are shown in Fig.~\ref{variables}, while the relevant data are 
summarized in Table~\ref{table:2}. 

We searched the literature and stellar catalogues in order to find prior 
variability information on the four objects. In particular, we looked at 
the works of \citet{Dam-ST07} and \citet{Dam-NR07}, and we searched 
the General Catalogue of Variable Stars (GCVS;~\url{http://www.sai.msu.su/groups/cluster/gcvs/gcvs/}), 
the New Catalogue of Suspected Variable Stars (\url{http://www.sai.msu.su/groups/cluster/gcvs/gcvs/nsv/}), 
the VSX and ASAS databases (\url{http://www.aavso.org/vsx/} and \url{http://www.astrouw.edu.pl/asas/?page=acvs}),
and the VizieR database (\url{http://vizier.u-strasbg.fr/}). We failed to identify counterparts 
of our objects, so that we can report them here as new discoveries. 

\begin{figure}[h]
   \centering
   \includegraphics[scale=.65]{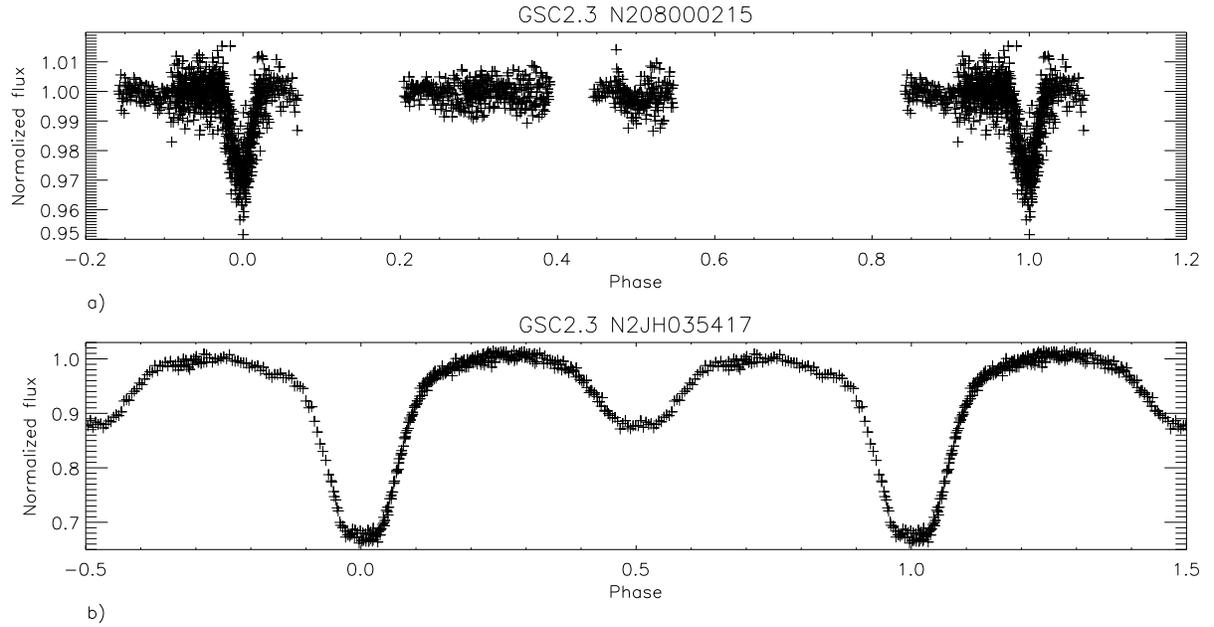}
      \caption{Phase-folded light curves of two eclipsing binary systems discovered 
      during the feasibility study. Upper panel: the estimated orbital period is 
      $P=2.02065\pm0.00004$ d and phase=0.0 corresponds to HJD=$2455041.50975 + P\times E$. 
      Lower panel: the estimated orbital period is $P=0.505167\pm0.000002$ d 
      and phase=0 corresponds to HJD=$2455031.381135 + P\times E$. In the upper and lower panels, there are
	  1510 and 1443 data points, respectively, taken over 11 and 6 days, respectively. }
         \label{bin1}
\end{figure}

The first two targets in Table~\ref{table:2} (GSC2.3 N208000215 and GSC2.3 N2JH035417) 
are identified as eclipsing binary systems, with orbital periods 
$P=2.02065\pm0.00004$ days and $P=0.505167\pm0.000002$ days, respectively, 
based on a Lomb-Scargle periodogram analysis. In the two panels of Fig.~\ref{bin1} we show 
the phase-folded light curves for these two systems. 
For the first binary system (upper panel of Fig.~\ref{bin1}) we derived the ephemeris of the secondary 
minimum, that was observed at the predicted epochs even if its depth is shallow 
($\sim$0.005 mag) and very close to the detection limit for our instrumentation.  

\begin{figure}[h]
   \centering
   \includegraphics[scale=.65]{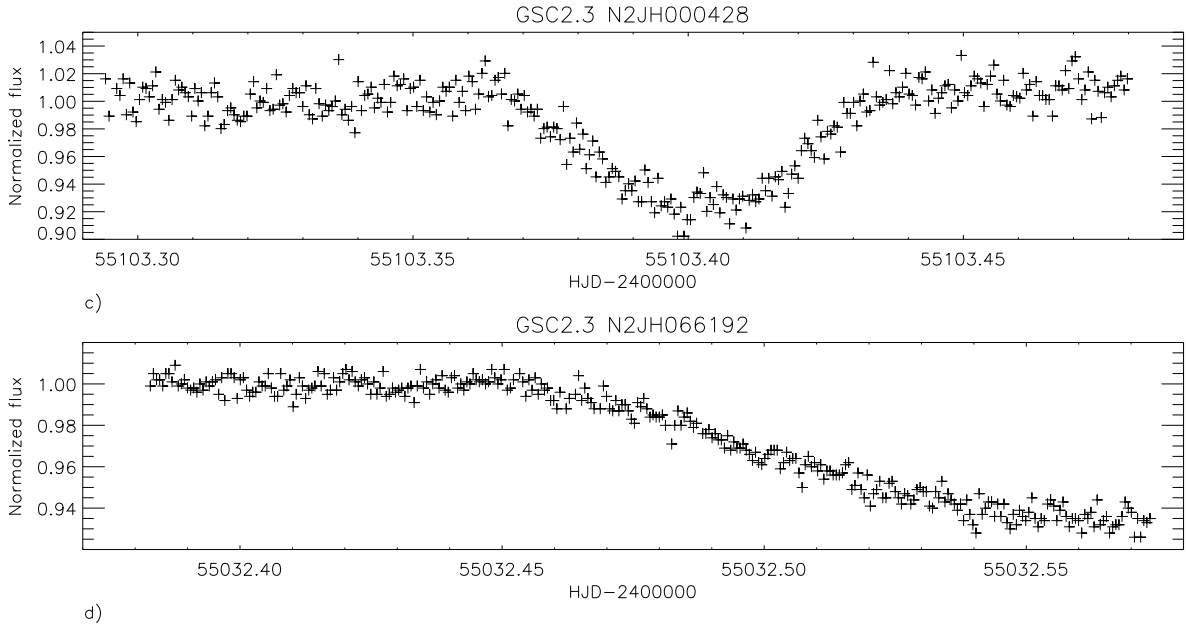}
      \caption{Light curves of the third and fourth object in Table~\ref{table:2} (field of the 
      star HAT-P-7) showing the occurrence of eclipses likely due  (based on 
      the observed depths) to a stellar companion. Both signals are seen only once in our data. 
      The minimum in the upper panel was detected during an 
      observing session aimed at gathering more data for the eclipsing binary system 
      showed in the lower panel of Fig.~\ref{bin1}. In the upper and lower panels, there are
	 317 and 313 data points, respectively, taken over a single night.}
         \label{bin2}
\end{figure}

For the last two targets in Table~\ref{table:2}, we observed in their light curves only one 
complete minimum (GSC2.3 N2JH000428) and a partial minimum (GSC2.3 N2JH066192), 
as showed in the upper and lower panels of Fig.~\ref{bin2} respectively. Because we observed 
these minima just one time, without detecting any other evident variability in other 
data, we can not estimate the orbital periods of the systems and we can only classify 
them generally as eclipsing binary stars.

\section{Concluding remarks}\label{concl}

We have carried out a site characterization campaign at the 
Astronomical Observatory of the Autonomous Region of the Aosta Valley (OAVdA), 
in order to demonstrate that OAVdA can be the location for a long-term 
photometric survey with small telescopes aimed at detecting transiting rocky planets 
around nearby cool dwarf stars. 

We focused the attention on those site-dependent factors that can have the largest impact 
on the ultimately achievable precision of the photometric measurements, such as 
seeing, extinction and night-sky brightness. We then correlated them with the quality of 
the photometric measurements of selected target fields, monitored with 20-40 cm 
class telescopes available on site, and analyzed using standard techniques of differential 
aperture photometry using an {\it ad hoc} developed data processing and analysis pipeline. 
Relevant data were collected in situ over a 
period of $\sim4$ months, and complemented by additional meteorological and photometric datasets 
covering a timespan of $> 3$ years. The main findings of this study (partly presented 
already in Damasso et al. 2009) can be summarized as follows: 

\begin{itemize}

\item[$\bullet$] The measured zenithal $V$-band night-sky brightness is typical of 
that of very good, very dark observing sites. The extinction registered at $V$ band is 
not dissimilar from that of other sites (for the same season). 

\item[$\bullet$] The median seeing over the period of in situ observations is found to be 
$\sim1.7^{\prime\prime}$. Given the limited duration of the observations, we did not probe 
any possible seeing seasonal patterns, or the details of its possible dependence on other 
meteorological parameters, such as wind speed and direction. Moreover, our data show that 
the seeing at the observatory was reasonably stable during most of the nights. 

\item[$\bullet$] The fraction of fully clear nights per year amounts to 39\%, while the total of 
useful nights increase to 57\% assuming a typical cloud cover of not more than 50\% of the night. 
There is not yet enough statistics to determine a clear value for the fraction of photometric nights per year. 

\item[$\bullet$] During 24 good observing nights over a period of over 3 months the median 
photometric precision was 0.006 mag for stars with magnitude $R\lesssim13$ mag. 
The typical nightly photometric precision appears to be uncorrelated with the seeing, 
whose typical value (in principle not competitive with other highly reputed observing sites 
around the world) turns out not to be a limiting factor for achievement of the photometric 
performance required for detection of planetary transits. 

\item[$\bullet$] The observation of known transiting planets (those in particular of 
WASP-3b and HAT-P-7b) allowed us to show that our differential aperture photometry approach 
applied to data collected with a small, 250-mm telescope, achieves best-case rms precision 
of $0.002-0.003$ mag for the brightest stars. 

\item[$\bullet$] A byproduct of the significant amount of photometric data collected 
in the stellar fields of WASP-3 and HAT-P-7 was the identification of a handful of new 
variable stars, four of which are presented and discussed here.

\end{itemize}

The results of the site testing campaign show conclusively that OAVdA is a suitable choice 
for hosting a long-term photometric survey for transiting planets around cool stars in the solar 
neighborhood.  A preliminary step towards the definition of such a long-term photometric transit search 
program is the 'pilot study' which we have begun earlier this year. We are devoting one of 
the 250 mm telescopes and the 400 mm instrument to the $I$-band photometric monitoring of $\sim50$ 
M dwarfs, within $\sim25$ pc from the Sun, with high-precision parallax estimates determined 
within the context of the TOPP program~\citep{Dam-Sma07, Dam-Sma10}. This study, 
with a targeted duration of less than 1 year, aims at obtaining uniform photometric 
monitoring of all targets (on the order of at least a few weeks of data per target), with the 
goal of providing benchmarks for the photometric micro-variability characteristics 
of M dwarfs as a function of their spectral sub-type (the stars in our sample span the range 
M0 through M9), in relation to other, primarily spectroscopic, activity indexes 
(e.g., based on H$_\alpha$, \ion{Ca}{2} lines emission). 
While the pilot study will likely not detect any planetary transits (given the 
limited number of objects in the target sample), the large photometric dataset collected 
will help us placing useful reference points for improved understanding of the impact 
on transiting planets detectability of stellar activity-related phenomena 
(e.g., spot-induced rotational modulation), and the prospects for taking them into 
account via detailed modeling procedures. 

In parallel to the science-driven activities of the pilot study, we are making progress 
in the re-organization of the scientific terrace for the purpose of hosting multiple 
identical 400-mm robotic telescope systems (at least five), the start of the operations 
of the first pair being envisaged in connection with the end of the pilot study (Fall 2010). 
The hardware and software requirements for the control system for data taking and storage 
have been clearly defined. Finally, as mentioned in \S~\ref{analysis}, necessary 
upgrades to the TEEPEE automated data reduction and analysis pipeline are being 
considered, in order to increase its functionality and robustness, ahead of the official 
start of our Western Italian Alps photometric transit search for planets around M dwarfs. 

\acknowledgements

An anonymous referee provided very useful, insightful comments which helped to materially improve the paper. 
MD, PC, and AB are supported by grants of the European Union, 
the Autonomous Region of the Aosta Valley and the Italian Department for Work, Health and Pensions.
The OAVdA is supported by the Regional Government of Valle d'Aosta, 
the Town Municipality of Nus and the Monte Emilius Community.
This research has been partially supported by INAF through PRIN 2009
``Environmental effects in the formation and evolution of extrasolar planetary system''.
AS gratefully acknowledges support from the Italian Space Agency (Contract ASI-Gaia I/037/08/0). 
We also thank A. Carbognani (OAVdA) 
and Davide Cenadelli (University of Milan) for useful discussions, D. Berger and ARPA-Valle d'Aosta for the collaboration 
in the study for the evaluation of the night-sky light pollution and the Servizio Centro 
Funzionale - Ufficio Meteorologico (Regione Autonoma Valle d'Aosta) which 
manages the weather station located next to OAVdA.
This research has made use of the SIMBAD database, operated at CDS, Strasbourg, France, 
and of NASAs Astrophysics Data System Bibliographic Services.



\newpage

\begin{sidewaystable}[h]
\caption{Summary of the main characteristics of the telescope systems described in \S~2.}             
\label{table:1}      
\centering                          
\begin{tabular}{c c c c c c c}        
\hline\hline                 
&\multicolumn{2}{c}{Telescope}&\multicolumn{2}{c}{CCD camera}&\multicolumn{2}{c}{Resulting configuration} \\  
\hline
Optical scheme&Aperture&Focal ratio&Sensor area&Pixel area&FoV&Plate scale\\
&(mm)& &(pixel)&($\mu$m)&(arcmin)&($^{\prime\prime}$/pixel)\\
\hline                   
Reflector Maksutov & 250 & f/3.80 & $2184\times1472$ & $6.8\times6.8$ & $52.10\times35.11$ & 1.43 (binning $1\times1$) \\
Reflector Ritchey-Chr\'{e}tien & 400 & f/7.64 & $1024\times1024$ & $24\times24$ & $26.4\times26.4$ & 1.55 (binning $1\times1$) \\
Reflector Ritchey-Chr\'{e}tien & 810 & f/7.90 & $2048\times2048$ & $15\times15$ & $16.3 \times16.3$ & 0.48 (binning $1\times1$)\\
\hline\hline                            
\end{tabular}
\end{sidewaystable}

\clearpage 

\begin{sidewaystable}[h]
\caption{Variable stars discovered in the fields of the extrasolar planet hosting 
stars WASP-3 and HAT-P-7. $R$ magnitudes are derived from $r^\prime$ magnitudes of the CMC14 
catalogue using the relation $R=r^\prime-0.22$ derived by \citet{Dam-DM09}} 
\label{table:2} 
\centering 
\begin{tabular}{c c c c c c c} 
\hline\hline 
Object ID & Eq. Coord. & & &Magnitude& & \\
\hline
(GSC2.3 catalogue)& RA, DEC (J2000) & B & V & R & J & K\\
\hline 
N208000215 & 18:35:23.210 +35:14:45.43 & 12.34$\pm$0.19 & 12.09$\pm$0.18 & 11.622 & 10.962$\pm$0.02 & 10.662$\pm$0.023\\ 
N2JH035417&19:27:49.596 +48:20:47.90 & - & - & 12.797 & 12.365$\pm$0.022 &  12.225$\pm$0.024  \\ 
N2JH000428 & 19:29:55.081 +47:50:04.41 & 13.34$\pm$0.30 & 14.10$\pm$0.44 & 12.95 & 11.998$\pm$0.020 &  11.053$\pm$0.019\\ 
N2JH066192 & 19:30:08.505 +47:49:31.16 & 10.11$\pm$0.04 & 8.92$\pm$0.02 &-& 6.857$\pm$0.018 & 6.205$\pm$0.016\\ 
\hline\hline 
\end{tabular}
\end{sidewaystable}

\clearpage

\newpage
 
\newpage 

\newpage

\newpage 

\newpage 

\newpage 

\newpage 

\newpage

\newpage 

\newpage 

\newpage 

\newpage 

\newpage 

\newpage 

\newpage 

\newpage 

\newpage 

\end{document}